\documentclass[12pt]{article} 
\usepackage{amsmath}
\usepackage{amssymb,amsfonts}
\usepackage{hyperref}
\renewcommand{\baselinestretch}{1.2}
\newcommand{\irrep}[1]{\ensuremath{\boldsymbol{#1}}}
\newcommand{\eis}[3]{~\ensuremath{{\cal E}^{#1}_{\irrep{#2};#3}}}
\newcommand{\exc}{\ensuremath{E_{d+1(d+1)}}}
\newcommand{\ie}{i.e.\ }

\newcommand{\Tr}{{\rm Tr}\ }

\newcommand{\C}{{\cal C}}
\newcommand{\E}{{\cal E}}

\newcommand{\K}{{\cal K}}
\newcommand{\M}{{\cal M}}
\newcommand{\R}{{\cal R}}
\renewcommand{\SS}{{\cal S}}
\newcommand{\T}{{\cal T}}
\newcommand{\F}{{\cal F}}

\newcommand{\Zint}{\mathbb{Z}}
\newcommand{\Real}{\mathbb{R}}
\newtheorem{theo}{Theorem}
\newtheorem{guess}[theo]{Conjecture}
\newtheorem{prop}[theo]{Proposition}
\newtheorem{corol}[theo]{Corollary}

\def\tr{\mbox{Tr}}
\def\pf{\mbox{Pfaffian}}

\def\hB{\hat{B}}
\def\tB{\tilde{B}}
\def\hg{\hat{g}}
\def\pa{\partial}
\newcommand{\vect}{\ensuremath{{\bf V}}}
\newcommand{\spi}{\ensuremath{{\bf S}}}
\newcommand{\spb}{\ensuremath{{\bf C}}}
\newcommand{\be}{\begin{equation}}
\newcommand{\ee}{\end{equation}}
\newcommand{\ben}{\begin{enumerate}}
\newcommand{\een}{\end{enumerate}}
\renewcommand{\sp}{\ ,\qquad}
\makeatletter
\renewcommand{\@makefnmark}{\mbox{$^{\ddagger\@thefnmark}$}}
\renewcommand{\subsection}{\@startsection
  {subsection}{2}{0pt
}{-\baselineskip}{0.5\baselineskip}
  {\normalfont\normalsize\itshape}}
\renewcommand{\section}{\@startsection
  {section}{2}{0pt
}{-\baselineskip}{0.5\baselineskip}
  {\bf\large}}

\makeatother
\numberwithin{equation}{section}
\numberwithin{table}{section}

\DeclareMathOperator{\rank}{rk}

\newcommand{\publititle}[8]
{ 
  \vspace*{-3cm}
  \begin{flushright} #1 \\ {\tt #2} \end{flushright}
  \vfill
  \begin{center}{\Large
    \bfseries #3}\end{center}
  \vskip 8mm
  \begin{center}{\large #4}\end{center}
  \begin{center}{\normalsize\sl #5}\end{center}
  \vskip 8mm
  \nopagebreak
  \noindent #6
  \vfill
  \begin{flushleft} #7
  \end{flushleft}
  \hrule width 6.7cm \vskip.1mm
  {\small #8}
  \thispagestyle{empty}
  \clearpage
}

\begin{document}

\publititle{NORDITA-1999/18 HE \\ NBI-HE-99-06\\CPHT--S710-0299 }
{hep-th/9903113v4} {Eisenstein Series and String Thresholds
${}^{\star}$} {N.A. Obers$^{a\dagger}$ and B.
Pioline$^{\,b\ddagger}$ } {$^a$Nordita and Niels Bohr Institute,
\\ Blegdamsvej 17,DK-2100 Copenhagen, Danmark\\[2mm] $^b$Centre de
Physique Th{\'e}orique, \\ Ecole Polytechnique$^\ast$, {}F-91128
Palaiseau, France } {We investigate the relevance of Eisenstein
series for representing certain $G(\Zint)$-invariant string theory
amplitudes which receive corrections from BPS states only.
$G(\Zint)$ may stand for any of the mapping class, T-duality and
U-duality groups $SL(d,\Zint)$, $SO(d,d,\Zint)$ or $\exc(\Zint)$
respectively. Using $G(\Zint)$-invariant mass formulae, we
construct invariant modular functions on the symmetric space
$K\backslash G(\Real)$ of non-compact type, with $K$ the maximal
compact subgroup of $G(\Real)$, that generalize the standard
non-holomorphic Eisenstein series arising in harmonic analysis on
the fundamental domain of the Poincar{\'e} upper half-plane.
Comparing the asymptotics and eigenvalues of the Eisenstein series
under second order differential operators with quantities arising
in one- and $g$-loop string amplitudes, we obtain a manifestly
T-duality invariant representation of the latter, conjecture their
non-perturbative U-duality invariant extension, and analyze the
resulting non-perturbative effects. This includes the $R^4$ and
$R^4 H^{4g-4}$ couplings in toroidal compactifications of M-theory
to any dimension $D\ge 4$ and $D\ge6$ respectively. }
{NORDITA-1999/18 HE,\ NBI-HE-99-06,\ CPHT--S710-0299,\\March 1999, revised Jan 2010.
\\ {\it Comm. Math. Phys. 209 (2000) 275-324
} } {$^{\ast}${\small Unit{\'e} mixte CNRS UMR 7644} \hfill
$^{\dagger}${\tt obers@nordita.dk} \hfill $^{\ddagger}${\tt
pioline@cpht.polytechnique.fr} \\ $^{\star}$ Work supported in
part by TMR networks ERBFMRXCT96-0045 and ERBFMRXCT96-0090.}
\clearpage

\renewcommand{\baselinestretch}{.9}\rm
\tableofcontents
\renewcommand{\baselinestretch}{1.2}\rm

\section{Introduction}
While worldsheet modular invariance has played a major role in the
context of perturbative string theory since its early days, the
advent of target space and non-perturbative dualities has brought
into play yet another branch of the mathematics of automorphic
forms invariant under infinite discrete groups. Indeed, physical
amplitudes should depend on scalar fields usually taking values
(in theories with many supersymmetries) in a symmetric space
$K\backslash G(\Real)$, where  $K$ is the maximal compact subgroup
of $G$, while duality identifies points in $K\backslash G(\Real)$
differing by the right action of an infinite discrete subgroup
$G(\Zint)$ of $G(\Real)$. This includes in particular the mapping
class group $SL(d,\Zint)$ in the case of toroidal
compactifications of diffeomorphism-invariant theories, the
T-duality group $SO(d,d,\Zint)$ in toroidal compactifications of
string theories, as well as the non-perturbative U-duality group
$\exc(\Zint)$ in maximally supersymmetric compactified M-theory
\cite{Hull:1995mz,Townsend:1995kk,Witten:1995ex} (see for instance
\cite{Giveon:1994fu,Obers:1998fb} for reviews and exhaustive list
of references). Moreover, supersymmetry constrains certain ``BPS
saturated'' amplitudes to be eigenmodes of second order
differential operators
\cite{Pioline:1998mn,Paban:1998qy,Green:1998by}, so that harmonic
analysis on such spaces provides a powerful tool for understanding
these quantities. In the most favorable case, it can be used to
determine exact non-perturbative results not obtainable otherwise,
which can then be analyzed at weak coupling
\cite{Green:1997tv,Kiritsis:1997em}. Other exact results can also
be obtained from string-string duality, although in a much less
general way, since one needs to be able to control the result on
one side of the duality map. This approach was taken for vacua
with 16 supersymmetries  in
\cite{Harvey:1996ir,Gregori:1997hi,Bachas:1997mc,Kiritsis:1997hf,Antoniadis:1997zt,Lerche:1998nx,Lerche:1998gz,Foerger:1998kw}.
In both cases, one generically obtains a few perturbative leading
terms which can in principle be checked against a loop
computation, whereas the non-perturbative contributions correspond
to instantonic saddle points of the unknown string field theory. A
number of hints for the rules of semi-classical calculus in string
theory have been extracted from these results
\cite{Bachas:1997xn,Green:1998tn}
 and reproduced
in Matrix models \cite{Moore:1998et,Kostov:1998pg,Gava:1999zk}, but a complete
prescription is still lacking.
A better understanding of such effects would be
very welcome, as it would for instance allow quantitative
computations of perturbatively-forbidden processes
in cases of more immediate physical relevance.

The prototypical example was proposed by Green and Gutperle, who
conjectured that the $R^4$ couplings in ten-dimensional type IIB
theory were exactly given by an S-duality invariant result
\cite{Green:1997tv}
\begin{equation}
\label{r4iib}
f_{R^4}^{IIB}= \frac{1}{ l_P^{2}}
\sum_{(m,n)\ne 0} \left[ \frac{\tau_2}{|m+n\tau|^2} \right]^{3/2}
=\frac{\zeta(3)}{l_P^{8}} \sum_{(p,q)=1} \frac{1}{\T^3_{(p,q)}}
\end{equation}
where in the first expression $\tau=a+i/g_s=\tau_1+i\tau_2$ is the
complexified string coupling transforming as a modular parameter
under $SL(2,\Zint)_S$ and $l_P=g_s^{1/4} l_s$ the S-duality
invariant ten-dimensional Planck length. This result is
interpreted in the second expression as a sum over the solitonic
$(p,q)$ strings of tension $\T_{(p,q)}=|p+q\tau|/l_s^2$. In
particular, the scaling dimension $-8+3\times 2$ is appropriate
for an $R^4$ coupling in ten dimensions. The invariant function in
\eqref{r4iib} is a particular case $s=3/2$ of a set of
non-holomorphic automorphic forms
\begin{equation}
\label{esl2}
\eis{SL(2,\Zint)}{\irrep{2}}{s}
=\sum_{(m,n)\ne 0} \left[ \frac{\tau_2}{|m+n\tau|^2} \right]^{s}
\end{equation}
also known as Eisenstein series, which together with a discrete
set of cusp forms generate the spectrum of the Laplacian on the
fundamental domain of the upper half-plane, within the set of
modular functions increasing at most polynomially as $\tau_2\to
\infty$ (see \cite{Terras:1985} for an elementary introduction):
\begin{equation}
\label{lsl2}
\Delta_{U(1)\backslash SL(2)} \eis{SL(2,\Zint)}{2}{s} = \frac{s(s-1)}{2}
\eis{SL(2,\Zint)}{2}{s} \ ,\qquad
\Delta_{U(1)\backslash SL(2)}=
\frac{1}{2}\tau_2^2(\partial_{\tau_1}^2+\partial_{\tau_2}^2)\ .
\end{equation}
Cusp forms are exponentially suppressed at large $\tau_2$,
and lie at discrete values along the $s=1/2+i\Real$ axis, although
no explicit form is known for them. Eisenstein series on the other
hand can be expanded at weak coupling (large $\tau_2$)
by Poisson resummation on the integer $m$ (see Appendix \ref{lve}
for useful formulae):
\begin{align}
\label{eisfour}
\eis{SL(2,\Zint)}{\irrep{2}}{s}
=&2\zeta(2s)\tau_2^s + 2\sqrt{\pi} \tau_2^{1-s}
\frac{\Gamma(s-1/2)}{\Gamma(s)} \zeta(2s-1) \nonumber\\
&+  \frac{2 \pi ^s \sqrt{\tau_2}}{\Gamma(s)}
\sum_{m\ne 0} \sum_{n\ne 0}
\left| \frac{m}{n} \right|^{s-1/2}
K_{s-1/2} \left(2\pi \tau_2 |mn|\right)
e^{2\pi i m n \tau_1}
\end{align}
For $s=3/2$, this exhibits a tree-level and one-loop term which
can be checked against a perturbative computation, together with
an infinite series of instantonic effects, from the saddle point expansion
\eqref{asbessel} of the modified Bessel function $K_{1}$:
\begin{equation}
\label{di}
\eis{SL(2,\Zint)}{\irrep{2}}{s=3/2}=
2\zeta(3) e ^{-3\phi/2} + \frac{2\pi ^2}{3} e ^{\phi/2} +
4\pi \sum_{N\ne 0} \sum_{n|N} \frac{ \sqrt{N} }{n^2}
\left[ e^{-2\pi N (e^{-\phi} + i a)}
+ e^{-2\pi N (e^{-\phi} - i a)} \right]
+\dots
\end{equation}
where $e ^\phi=g_s$ denotes the type IIB coupling. These effects
can be interpreted as arising from D-instantons and anti-D-instantons
\cite{Green:1997tv}. As suggested in \cite{Green:1997di},
one can in fact prove that the 32 supersymmetries of type IIB
imply that the exact $R^4$ coupling should be an eigenmode of the
Laplacian on the moduli space $U(1)\backslash SL(2,\Real)$, with a definite
eigenvalue (3/8 in the conventions of the present paper)
\cite{Berkovits:1997pj,Pioline:1998mn,Green:1998by},
which uniquely selects out the $s=3/2$ Eisenstein series.
In particular, it rules out contributions from cusp forms,
which on the basis of the leading perturbative terms alone would
have been acceptable \cite{Pioline:1998mn}.

Whereas harmonic analysis on the fundamental domain of the
upper half-plane $U(1)\backslash$ $SL(2,\Real)/SL(2,\Zint)$ is
rather well understood, it is not so for the more
general symmetric spaces of interest in string theory
(see however \cite{Terras:1985ii,Harish:1968}).
It is the purpose of this work to generalize these considerations
to more elaborate cases, corresponding to a larger moduli space
and discrete symmetry group. Such situations arise
in compactifications with $16$ or $32$ supersymmetries, where
supersymmetry prevents corrections to the scalar manifold.
As we mentioned, this is the case of toroidal
compactifications of string theories, with (part of the)
moduli space $[SO(d)\times SO(d)]\backslash SO(d,d,\Real)/SO(d,d,\Zint)$,
or of M-theory, with moduli space $K\backslash\exc(\Real)/\exc(\Zint)$.
This also happens in
more complicated cases, such as type IIA on $K_3$, with
moduli space $\Real^+ \times [SO(4)\times SO(20)\backslash SO(4,20,\Real)]$
identified by the $SO(4,20,\Zint)$ (perturbative) mirror symmetry, or type IIB
on $K_3$, with moduli space $[SO(5)\times SO(21)] \backslash SO(5,21,\Real)$
identified by the non-perturbative symmetry $SO(5,21,\Zint)$.
It is even possible to have uncorrected tree-level scalar manifolds
in theories with $8$ supersymmetries, as in the FHSV model
\cite{Ferrara:1995yx}, with a moduli space $K\backslash[SL(2,\Real)\times
SO(2,10,\Real) \times SO(4,12,\Real)]$ where $K$ is the obvious maximal compact
subgroup. Note that in this case the duality group is broken to a
subgroup of $SO(2,10,\Zint) \times SO(4,12,\Zint)$, due to the effect
of the freely acting orbifold. Usually however,
in cases with 8 supersymmetries, amplitudes are
given by sections of a symplectic
bundle on the corrected moduli space, and our methods will not carry
over in a straightforward way.

This generalization was in fact started in Ref.
\cite{Kiritsis:1997em}, where toroidal compactifications of type
IIB string theory down to 7 or 8 dimensions  were considered. It
was demonstrated there that the straightforward extension of the
order 3/2 Eisenstein series (\ref{esl2}) to the U-duality groups
$SL(5,\Zint)$ and $SL(3,\Zint)$ reproduces the tree-level and
one-loop $R^4$ thresholds, together with $(p,q)$-string
instantons. A generalization to lower dimensional cases was also
proposed, and a distinct route using successive T-dualities was
taken in Ref. \cite{Pioline:1997pu} to obtain the contribution of
the $O(e^{-1/g_s})$ D-brane instantons in the toroidally
compactified type IIA and IIB theories; it was also pointed out
that S-duality suggests extra $O(e^{-1/g_s^2})$ contributions yet
to be understood. In the present work, we will take a more general
approach, and investigate the properties and utility of the
generalized Eisenstein series, that we define for any symmetric space
$K \backslash G(\Real)$ and any representation $\R$ of $G$ by
\begin{equation}
\label{geneis}
\eis{G(\Zint)}{\R}{s}(g) = \sum_{m\in \Lambda_\R\backslash\{0\} }
\delta( m\wedge m) ~\left[
m \cdot \R^t \R(g) \cdot m \right ]^{-s}\ .
\end{equation}
Here, $g$ denotes an element in the coset $K\backslash G(\Real)$,
$m$ a vector in an integer lattice $\Lambda_\R$ transforming in
the representation $\R$. $\wedge$ is an integer-valued product on
the lattice, such that the condition $m\wedge m=0$ projects the
symmetric tensor product $\R\otimes_s\R$ onto its highest
irreducible component, thus keeping only the ``completely
symmetric'' part. This definition is to be contrasted with the one
used in the mathematical literature \cite{Harish:1968}:
\begin{equation}
\label{geneis2}
{\cal E}^{G(\Zint)}_{\left\{w_i\right\} }(g) =
\sum_{h\in G(\Zint)/N}
\prod_{i=1}^r  a_i( gh )^{-w_i}
\end{equation}
where $w_i$ is now an arbitrary $r$-dimensional vector in weight
space, and $a(g)$ is the Abelian component of $g$ in the Iwasawa
decomposition of the rank-$r$ non-compact group $G(\Real)=K \cdot
A \cdot N$ into maximal compact $K$, Abelian $A$ and nilpotent $N$
subgroups. Note that this definition is manifestly $K$-invariant
on the left and $G(\Zint)$-invariant on the right. Choosing $w$
along a highest-weight vector $\lambda_{\R}$ associated to a
representation $\R$ reduces \eqref{geneis2} to \eqref{geneis}
where $w=s\lambda_{\R}$, up to an $s$-dependent factor. This
generalizes the equality in \eqref{r4iib} to higher rank groups.
The definition (\ref{geneis}), albeit less general, has a clearer
physical meaning: the lattice $\Lambda_\R$ labels the set of BPS
states in the representation $\R$ of the duality group, $\M^2=m
\cdot \R^t \R(g) \cdot m $ gives their mass squared (or tension),
and $m\wedge m=0$ imposes the half-BPS condition; this will be
shown to be a necessary requirement for the eigenmode condition
$\Delta_{K\backslash G} \eis{G(\Zint)}{\R}{s} \propto
\eis{G(\Zint)}{\R}{s}$, but could be dropped if one were to
address non--half-BPS saturated amplitudes.

The outline of this paper is as follows. In Section 2, we will discuss
the simplest case of $SL(d,\Zint)$ Eisenstein series, where most
of the features arise without the complications in the parametrization
of the moduli space. In Section 3, we will turn to $SO(d,d,\Zint)$
Eisenstein series, and discuss their applications for the computation
of T-duality invariant one-loop thresholds of string theories compactified
on a torus $T^d$. In Section 4, we covariantize this expression
to obtain exact non-perturbative $R^4$ couplings in
toroidal compactifications of M-theory to $D\ge 4$.
In Section 5, we apply the same techniques to the $g$-loop threshold,
and use it to deduce $R^4 H^{4g-4}$ exact couplings in
the same theory. Computational details will be relegated to the Appendices.
This work appeared on the archive simultaneously with
Ref. \cite{Ganor:1999}, which uses similar techniques, albeit with a different
motivation.

\section[Toroidal compactification and $SL(d,\Zint)$ Eisenstein series]
{Toroidal compactification and $SL(d,\Zint)$ Eisenstein series}

\subsection{Moduli space and Iwasawa gauge}
Infinite discrete symmetries appear in the simplest setting in
compactifications of a diffeomorphism invariant field theory on a
torus $T^d$. Specifying the internal manifold requires a flat
metric on the torus, that is a positive definite metric $g$.
Equivalently we may specify a vielbein $e\in Gl(d,\Real)$ such
that $g=e ^t e$, defined up to orthogonal rotations $SO(d,\Real)$
acting on the left, which leave $g$ invariant. This gauge
invariance can be fixed thanks to the Iwasawa decomposition
\begin{equation}
Gl(d,\Real)=SO(d,\Real)\times (\Real^+)^d \times \ N_d,
\end{equation}
where $(\Real^+)^d$ denotes the Abelian group of diagonal $d
\times d$ matrices with positive non zero entries, and $N_d$ the
nilpotent group of upper triangular matrices with unit diagonal,
by choosing $e$ in the last two factors, \ie in an upper
triangular form. The Abelian part corresponds to the radii of the
torus, whereas $N_d$ parametrizes the Wilson lines $A_i^j$  of the
Kaluza--Klein gauge field $g_{\mu i}$.

By general covariance, the Kaluza--Klein reduction of the field
theory on the torus only involves contractions with the metric
$g_{ij}$, so that the reduced theory is invariant under a symmetry
$h\in Gl(d,\Real)$ which transforms $g$ in the representation
$g\rightarrow h^t g h$. The vielbein on the other hand is acted
upon on the right, $e\to e h$, which has to be compensated by a
field-dependent $SO(d,\Real)$ gauge transformation $e\to
\omega(e,h) e$ to preserve the upper triangular form. Transforming
by an element $h\propto 1$ in the center of $Gl(d,\Real)$
corresponds to changing the volume, whereas an $SL(d,\Real)$
transformation affects the  torus shape. This change is not always
physical however, since an $SL(d,\Zint)$ rotation can be
compensated by a global diffeomorphism of the torus, \ie an
element of the mapping class group. The toroidal compactification
is therefore parametrized by the symmetric space
\begin{equation}
\label{slm}
\Real ^+ \times \left[ SO(d,\Real) \backslash SL(d,\Real) /
SL(d,\Zint) \right]
\end{equation}
and all physical amplitudes should be invariant under $SL(d,\Zint)$.
In particular, the effective action including the massive Kaluza--Klein
modes will only be invariant under $SL(d,\Zint)$, and not
$Gl(d,\Real)$.

\subsection{Fundamental and antifundamental Eisenstein series}
Keeping the above in mind, it is now straightforward to generalize the
$SL(2,\Zint)$ Eisenstein series \eqref{esl2} to the fundamental representation
of $SL(d,\Zint)$ as
\begin{equation}
\label{slef}
\eis{SL(d,\Zint)}{\irrep{d}}{s} = \hat{\sum_{m^i} }
\left[ m^i g_{ij} m^j \right]^{-s}\ ,
\end{equation}
where the subscript $\irrep{d}$ stands for the representation in
which the integers $m^i$, $i=1 \ldots d$ transform. In fact, the
above form is really a $Gl(d,\Zint)$ Eisenstein series since we
did not restrict $g$ to have unit determinant, but the dependence
on $V_d= \sqrt{\det g}$ is trivial so we shall keep with this
abuse of language. The $SL(d,\Zint)$-invariant form in
\eqref{slef} is easily seen to be an eigenmode of the Laplacian
\footnote{In all expressions for the Laplacians in the main text
of the paper we employ the convention that $\pa/\pa \tilde g_{ij}$
is taken with respect to the diagonally rescaled metric $\tilde
g_{ij} = (1 - \delta_{ij}/2 )g_{ij}$, and for simplicity of
notation we omit the tilde. As explained in Appendix  \ref{lap},
this redefinition has the advantage that unrestricted sums can be
used.} on the scalar manifold (\ref{slm}):
\begin{subequations}
\begin{equation}
\label{glfun}
\Delta_{Gl(d)} \eis{SL(d,\Zint)}{\irrep{d}}{s}=\frac{s(2s-d+1)}{2}
\eis{SL(d,\Zint)}{\irrep{d}}{s}
\end{equation}
\begin{equation}
\label{gllap}
\Delta_{Gl(d)}=
\frac{1}{4} 
g_{ik} g_{jl}
\frac{\partial}{\partial g_{ij}}
\frac{\partial}{\partial g_{kl}}
+ \frac{d+1}{4} 
g_{ij}
\frac{\partial}{\partial g_{ij}}
\end{equation}
\end{subequations}
In fact, it is more appropriate to restrict to the
$SO(d,\Real)\backslash SL(d,\Real)$ moduli, in terms of which
\begin{subequations}
\begin{equation}
\label{esllap}
\Delta_{SL(d)} \eis{SL(d,\Zint)}{\irrep{d}}{s}
=\frac{s(d-1)(2s-d)}{2d}
\eis{SL(d,\Zint)}{\irrep{d}}{s}
\end{equation}
\begin{equation}
\label{sllap}
\Delta_{SL(d)}=
\frac{1}{4} 
g_{ik} g_{jl}
\frac{\partial}{\partial g_{ij}}
\frac{\partial}{\partial g_{kl}}
-\frac{1}{4d} \left( 
g_{ij}\frac{\partial}{\partial g_{ij}} \right)^2
+\frac{d+1}{4} 
g_{ij}
\frac{\partial}{\partial g_{ij}}
\end{equation}
\end{subequations}

Here we may wonder why we should choose the integers $m$ to lie in
the fundamental representation $\irrep{d}$ of $SL(d)$. Choosing
$m$ to transform in the contragredient representation,\footnote{We denote the contragredient 
representations of $\R$ by $\bar \R$, not to be confused by complex conjugation (all finite dimensional representations considered in this paper are real).}
\begin{equation}
\label{sleaf}
\eis{SL(d,\Zint)}{\irrep{\bar{d}}}{s} = \hat{\sum_{m_i} }
\left[ m_i g^{ij} m_j \right]^{-s}\ ,
\end{equation}
does not bring much novelty, since a Poisson resummation
over all integers $m_i$ brings us back to Eq. (\ref{slef}),
albeit with a transformed order $s\to d/2-s$:
\begin{equation}
\label{slinv}
\eis{SL(d,\Zint)}{\irrep{\bar{d}}}{s}=
\frac{V_d \pi^{s} \Gamma(\frac{d}{2}-s)}{\pi^{\frac{d}{2}-s} \Gamma(s) }
\eis{SL(d,\Zint)}{\irrep{d}}{\frac{d}{2}-s}
\end{equation}
Note that the two series $\eis{SL(d,\Zint)}{\irrep{d}}{s}$ and
$\eis{SL(d,\Zint)}{\irrep{\bar{d}}}{s}$ have
the same eigenvalue under $\Delta_{SL(d)}$, but
different eigenvalues under  $\Delta_{Gl(d)}$. This simply stems
from their different dependence on the volume, and is not sufficient
to lift their degeneracy under the $SL(d)$ Laplacian.

\subsection{Higher representations and constrained Eisenstein series}
We may also choose $m$ to transform in a higher dimensional
representation, \ie as a tensor $m_{ij\dots}$ with prescribed
symmetry properties. In order to determine whether we still
get an eigenmode, it is useful to take a more algebraic approach.
We consider acting with the Laplacian on the integral representation
\begin{equation}
\label{intrep}
\left[m^t M m\right]^{-s}=\frac{\pi ^s}{\Gamma(s)}
\int_{0}^{\infty} \frac{dt}{t^{1+s}}
\exp \left( -\frac{\pi} {t} m^t M m
\right)
\end{equation}
where $M=\R^t\R$ denotes the mass matrix in the representation
$\R$. Deriving only once in the exponential yields the action of
the Laplacian on $M$, which transforms as a symmetric tensor
product $\R\otimes_s\R$. In order to get an eigenmode, this tensor
product should be irreducible when contracted with the charges
$m$. This puts a quadratic constraint on $m$, which we generically
denote $m\wedge m=0$. In other words, $m\wedge m=0$ projects onto
the highest irreducible component of the symmetric tensor product
$\R\otimes_s\R$. One may want to drop the quadratic constraint,
and still impose higher cubic and quartic constraints, in order to
obtain candidates for quarter-BPS amplitudes, but we will not
pursue this line here. Assuming this constraint is fulfilled, we
therefore get an insertion of $-\pi Q[\R\otimes \R]/4t$ in the integral,
where $Q[\SS]$ is the Casimir $(T^i)^2$ in the representation
$\SS$ (we normalize the Laplacian such that $\Delta=\frac14 (T^i)^2$). 
The other term with two derivatives acting in the
exponential gives a contribution $\pi^2\tilde Q[\R \otimes _s \R]/4t^2$,
where $\tilde Q[\SS]$ denotes the operator $T^i\otimes T^i$ acting
on the symmetric tensor product $\SS\otimes_s \SS$. By developing
the square in $(T^i\otimes 1+ 1\otimes T^i)^2$, we find that
$Q[\SS \otimes_s \SS]=2 Q[\SS] + 2 \tilde Q[\SS]$, so that all in
all
\begin{equation}
\label{delexp}
e^{\pi m^t M m/t}\Delta  e^{-\pi m^t M m/t} =\frac{\pi^2}{8t^2} \left(
Q[\R^{\otimes_s 4}] - 2  Q[\R^{\otimes_s}] \right)\,  (m^t M m)^2
- \frac{\pi}{4t}\, Q[\R^{\otimes_s 2}] \, (m^t M m)
\end{equation}
We now use the expression
for the Casimir of the $p$-th symmetric power of a representation
of highest weight $\lambda$,
\begin{equation}
\label{casi}
Q[\R_\lambda ^{\otimes_s p}] =( p \lambda,p\lambda+2\rho)
\end{equation}
where $\rho$ is the Weyl vector, \ie the sum of all the fundamental
weights, and $(\cdot,\cdot)$ the inner product on the weight space
with the length of the roots normalized to 2 (since we restrict
to simply laced Lie groups of $ADE$ type).
Using formula \eqref{ide} to integrate by part in \eqref{intrep}, we thus find
\begin{prop}
\label{vectheo}
The constrained Eisenstein series \eqref{geneis} associated to the
representation of highest weight $\lambda$ is an eigenmode of
the Laplacian with eigenvalue
\begin{equation}
\label{eisgen} \Delta_{K\backslash G}
\eis{G(\Zint)}{\irrep{\R_\lambda}}{s} = s (\lambda,s \lambda-\rho)
\eis{G(\Zint)}{\irrep{\R_\lambda}}{s}
\end{equation}
\end{prop}
This result reproduces the eigenvalue (\ref{esllap}) for the fundamental
representation of $SL(d,\Zint)$ but will be applied for many other
situations in the following. It implies in particular
that Eisenstein series associated to representations related
by outer automorphisms, \ie symmetries of the Dynkin diagram, are
degenerate under $\Delta_{K\backslash G}$, as well as two
Eisenstein series of same representation but order $s$
and $[(\lambda,\rho)/(\lambda,\lambda)]-s$. We also note that
\eqref{eisgen}
can be obtained more quickly by noting that $\M^{-2s}=(m^t Mm)^{-s}$
transforms as the symmetric power of order $-2s$ of $\R$, and
substituting $p=-2s$ in \eqref{casi}. Finally, we note that
Eisenstein series are in fact eigenmodes of the complete
algebra of invariant differential operators \cite{Harish:1968}.

In some cases, it may happen that the constraint $m \wedge m =0$
can be solved in terms of a lower dimensional representation. This
is for instance the case of $p$-th symmetric tensors of
$SL(d,\Real)$, where the constraint implies that the integers
$m^{ijkl\dots}$ themselves are, up to an integer $r$, the
symmetric power of a fundamental representation $n^i$:
\begin{equation}
m^{ijkl\dots} = r~ n^i n^j n^k n^l\dots\ .
\end{equation}
The summation over $r$ can then be carried out explicitly, and the
result is proportional to the Eisenstein function in the
fundamental representation, with a redefined order $s\to p s$.
This, however, does not happen for antisymmetric tensors. Since
the antisymmetric representations are associated to the nodes of
the Dynkin diagram, we thus see that a subset of eigenmodes of the                       
Laplacian is in general provided by                                                     
the Eisenstein series associated to the nodes of the Dynkin diagram,                    
up to cusp forms. 

\subsection{Decompactification and analyticity}
Our definition of Eisenstein series has so far remained rather
formal: the infinite sums appearing in (\ref{slef}), (\ref{sleaf})
are absolutely convergent for $s>d/2$ only, and need to be
analytically continued for other values of $s$ in the complex
plane\footnote{Other regularization methods have also been
discussed in Ref. \cite{Kiritsis:1997em}}. It turns out that the
analyticity properties can be determined by induction on $d$,
which corresponds to the physical process of decompactification.
We thus assume the torus $T^{d+1}$ to factorize into  a circle of
radius $R$ times a torus $T^d$ with metric $g_{ab}$ and use the
integral representation \eqref{intrep} of the Eisenstein series,
say in the fundamental representation,
\begin{equation}
\eis{SL(d+1,\Zint)}{d+1}{s}=\frac{\pi ^s}{\Gamma(s)}
\int_{0}^{\infty} \frac{dt}{t^{1+s}}
\hat{\sum_{m,n^a} } \exp \left( -\frac{\pi} {t}
\left[ n^a g_{ab} n^b  + R^2 m^2 \right]
\right)
\end{equation}
where $m$ denotes the first component of $n^i$ . The leading term
as $R\to\infty$ corresponds to the $m=0$ contribution, which
reduces to the $SL(d,\Zint)$ Eisenstein series. Subleading
contributions arise by Poisson resumming on the unrestricted (since
now $m\ne 0$) integers $n_a$:
\begin{equation}
\eis{SL(d,\Zint)}{d+1}{s}=\eis{SL(d,\Zint)}{d}{s} + \frac{\pi
^s}{V_d\Gamma(s)} \int_{0}^{\infty} \frac{dt}{t^{1+s-\frac{d}{2}}}
\sum_{n_a}\hat{\sum_m} \exp \left( -\pi t n_a g^{ab} n_b  -\frac{\pi} {t}
R^2 m^2 \right)
\end{equation}
where $V_d$ stands for the volume $\sqrt{\det g}$
of the torus $T^d$.
Separating the $n_a=0$ contribution from the still subleading
$n_a\ne 0$ one, we get
\begin{align}
\eis{SL(d+1,\Zint)}{d+1}{s} &= \eis{SL(d,\Zint)}{d}{s}
+ \frac{ 2 \pi^s \Gamma(s-d/2) \zeta(2s-d)}
{ \pi^{s-d/2} \Gamma(s) R^{2s-d} V_d} + \nonumber\\
&\frac{2\pi ^s}{\Gamma(s)R^{2s-d-2}}
\hat{\sum_m} \hat{\sum_{n_a}} \left| \frac{ n_a g^{ab} n_b}{m^2} \right|
K_{s-d/2} \left( 2 \pi |m| R \sqrt{  n_a g^{ab} n_b} \right)
\label{expsl}
\end{align}
Using the asymptotic behaviour of the Bessel function \eqref{asbessel}, we
see that the last term is exponentially suppressed of order
$O(e ^{-R})$, and the sum is absolutely convergent and thus
analytic in $s$. For $d=1$, the $SL(d,\Zint)$ Eisenstein series reduces
to $2 \zeta(2s) R^{-2s}$ and has a simple pole at $s=1/2$.
For $d>1$, induction shows that the pole at $s=d/2$ from the
second term cancels the one in the $SL(d,\Zint)$ Eisenstein
series, leaving the pole at $s=(d+1)/2$ from the zeta function
in (\ref{expsl}). We thus have
\begin{prop}
The $SL(d,\Zint)$ Eisenstein series of order $s$ in the fundamental
representation can be analytically continued to the $s$-plane with
$s=d/2$ excluded, where it has a single pole with residue
\begin{equation}
\label{ana}
\eis{SL(d,\Zint)}{d}{s} \simeq \frac{ \pi^{d/2} }{V_d \Gamma(d/2)}
\frac{1}{s-d/2}
\end{equation}
\end{prop}
This result is well known in the mathematical literature
\cite{Terras:1985ii}.  Of course, the same holds for the
antifundamental representation by replacing $V_d$ by its inverse.
Let us mention in passing that, together with the functional
relation (\ref{slinv}), this implies a relation which generalizes
$2\zeta(0)=-1$:
\begin{equation}
\eis{SL(d,\Zint)}{d}{s=0}=\eis{SL(d,\Zint)}{\bar{d}}{s=0}=-1
\end{equation}
We also note that the pole at $s=d/2$ coincides with the vanishing
of the eigenvalue of the Eisenstein series under the Laplacian
$\Delta_{SL(d)}$. This is so because the residue is moduli
independent. An invariant modular form can still be obtained
by subtracting the pole, in which case the eigenmode equation
gets a harmonic anomaly:
\begin{equation}
\label{harmanom}
\Delta_{SL(d)} ~\hat{\cal E}^{SL(d,\Zint)}_{\irrep{d};s=d/2} =
\frac{ \pi^{d/2} (d-1)}{2  \Gamma(d/2) V_d}
\end{equation}
The case $d=2$, particularly relevant in the sequel, corresponds to 
Kronecker's first limit formula (see e.g. \cite{Terras:1985}),
\begin{equation}
\label{ded} \hat{\cal E}^{SL(2,\Zint)}_{\irrep{2};s=1}
   = -\pi \log
\left(4 e^{-4\gamma} \tau_2 |\eta(\tau)|^4 \right)
\end{equation}
where $\eta(\tau)$ denotes the usual Dedekind function and $\gamma$ is Euler's constant.

This computation can unfortunately not be made for constrained
Eisenstein series, since the constraint prevents a simple
Poisson resummation. We shall come back to this problem in the
next section for the $SO(d,d,\Zint)$ case.
We can however conjecture the analytic
structure from a simple argument: the divergences arise from the
large $m$ region, where the integers can be approximated by $N=\dim
\R$ continuous
variables. The $N_c$ quadratic constraints
restrict the phase space to $\Real^{N-N_c}$, while inserting
an extra factor $r^{-N_c}$ in spherical coordinates, from
$\delta(r^2)=\delta(r)/2r$. We are therefore led to the integral
$\int r^{-N_c} r^{N-N_c-1} r^{-2s} dr$, which converges for
$s>(N-2N_c)/2$. We therefore expect a simple pole at
$s=(N-2N_c)/2$ for an Eisenstein series of an $N$-dimensional
representation with $N_c$ independent constraints.

\subsection{Partial Iwasawa decomposition}
In determining the decompactification behaviour, we assumed the
torus $T^{d+1}$ to factorize into $T^d\times S_1$. This may be
too restrictive, as for instance in M-theory applications, where
we are interested in the perturbative type II limit corresponding
to a vanishingly small circle of radius $R_s=g_s l_s$ but still
want to retain the effect of the off-diagonal metric, \ie the
Ramond one-form $\mathcal{A}$. It is then convenient to take the
Kaluza--Klein ansatz
\begin{equation}
\label{kkans}
dx^i g_{ij} dx^j  = R^2 (dx^1 + A_a d x ^a)^2 + d x^a \hg_{ab} d x^b
\end{equation}
which is nothing but a partial Iwasawa decomposition. This
breaks the higher dimensional symmetry $SL(d+1,\Real)$ to a
subgroup $SL(d,\Real)$, together with a nilpotent group of
constant shifts $A_a \to A_a + \Lambda_a$, which is what remains
from the Kaluza--Klein gauge invariance on a flat torus. In terms
of these variables, the Laplacian takes the form (See
Appendix \ref{dec} for details on the derivation.)
\begin{equation}
\label{gllapd}
\Delta_{Gl(d+1)}=\Delta_{Gl(d)} - \frac{1}{4}
\hg_{ab} \frac{\partial}{\partial \hg_{ab}}+
\frac{1}{4} \left( R\frac{\partial}{\partial R} \right)^2
+ \frac{d}{4}  R\frac{\partial}{\partial R}
+ \frac{\hg_{ab}}{2R^2}
\frac{\partial}{\partial A_a}\frac{\partial}{\partial A_b}
\end{equation}
One can then check that each term in (\ref{expsl}) -- upon
reinstating the dependence on $A_a$ -- is an eigenmode of
the Laplacian with the correct eigenvalue.

\section{$SO(d,d,\Zint)$ Eisenstein series and one-loop thresholds}

In this section we turn to the construction of Eisenstein series
for $SO(d,d,\Zint)$ and its application to one-loop thresholds in
type II string theory. Higher genus contributions are also
amenable to an Eisenstein series representation, and will be
addressed in Section \ref{highg}.

\subsection{Moduli space and Iwasawa gauge}
Owing to the occurrence of winding states
charged under the 2-form $B_{\mu\nu}$,
any closed string theory on a torus
$T^d$ exhibits a larger symmetry $O(d,d,\Zint)$,
a discrete subgroup of the $O(d,d,\Real)$ symmetry
of the massless degrees of freedom. The symmetry is
actually reduced to $SO(d,d,\Zint)$ in type II theories,
where the elements in $O(d,d,\Zint)$ with determinant $-1$
map type IIA to type IIB. This T-duality
is valid to all orders in perturbation theory, and
postulated to hold non-perturbatively as well. It
contains the mapping class group $SL(d,\Zint)$ of the torus
as a subgroup, as well as generators that are
non-perturbative from a world-sheet
point of view. The moduli space
includes a symmetric subspace
\begin{equation}
\label{som}
 \left[ SO(d,\Real) \times SO(d,\Real)\right] \backslash SO(d,d,\Real) /
SO(d,d,\Zint)
\end{equation}
describing the metric of the torus and the two-form background,
which can again be parametrized using the Iwasawa decomposition
\begin{equation}
SO(d,d,\Real)=[SO(d,\Real) \times SO(d,\Real)] \times (\Real^+)^{d} \times
\ N_{2d}^{SO},
\end{equation}
More precisely, the Abelian part $(\Real^+)^{d} $ corresponds to
the $d$ radii (and $d$ inverse radii) of the torus and the
nilpotent part $N_{2d}^{SO}$ parametrizes the Wilson lines $A_i^j$
of the Kaluza--Klein gauge field and the antisymmetric tensor
$B_{ij}$. In particular, in the basis where the $SO(d,d,\Zint)$
invariant tensor is $\eta=\scriptstyle
\begin{pmatrix} 0&1_d\\1_d&0\end{pmatrix}$, the gauge-fixed vielbein
$e$  can be chosen as
\begin{equation}
\begin{tiny} \label{iwast} e = \left( \begin{array}{ccc|ccc} 1/R_1 & & & &
& \\
    & 1/R_2    &       &       &    & \\
    &        &\ddots &       &    & \\
\hline
    &        &       & R_1   &    & \\
    &        &       &       & R_2 & \\
    &        &       &       &    &\ddots
\end{array} \right)
\cdot \left( \begin{array}{ccc|ccc} 1 &    & \dots & & &\\
    -A_2^1 & 1      & \ddots& &  & \\
    &        & \ddots& &  &     \\
\hline
    &        &       & 1     & A_2^1 &\dots\\
    &        &       &       & 1      &\ddots \\
    &        &       &       &        &\ddots \\
\end{array} \right)
\cdot  \left( \begin{array}{ccc|ccc}  1 &     & &B_{11} & B_{12}
&\dots\\
    &  1    &  &B_{21} & B_{22} &\dots\\
    &        & \ddots &\vdots & \vdots &     \\
\hline
     &        &       & 1      &  & \\
    &          &       &       & 1       & \\
    &        &       &       &        & \ddots \\
\end{array} \right)
\end{tiny}
\end{equation}
and right symmetry transformations by an $SO(d,d,\Real)$ element
have to be compensated by left $SO(d,\Real)\times SO(d,\Real)$
gauge transformations. The $SL(d,\Real)$ subgroup corresponds to
block diagonal elements $\scriptstyle\begin{pmatrix}
g^{-1}&\\&g\end{pmatrix}$. In analogy with the $SO(d)\backslash
SL(d,\Real)$ case, we can trade the vielbein $e$ for the gauge
invariant moduli matrix
\begin{equation}
\label{modmat}
M(\irrep{\vect}) = e^t e = \left(
\begin{array}{cc}
g^{-1} & g^{-1} B \\ -B g^{-1}  & g- B g^{-1} B  \\
\end{array}
 \right)
\end{equation}
which provides the mass matrix for BPS states in the vector
representation $\vect$ of $SO(d,d)$, namely momentum and winding states.
D-branes on the other hand transform as (conjugate) spinor representations
$\spi$ ($\spb$),
and their mass matrix is given accordingly by $\R(e)^t \R(e)$
where $\R(e)$ is the spinor or conjugate spinor representation
of the group element $e$. We can therefore build $SO(d,d,\Zint)$
invariant functions by summing the BPS mass or tension
over all BPS states, which we do now.

\subsection{Spinor and vector Eisenstein series}
In order to define these T-duality invariant functions, we need
to be more explicit about the mass matrix and BPS conditions of
these states. These have been reviewed in \cite{Obers:1998fb}
(see \cite{Obers:1998rn} for a r{\'e}sum{\'e})
so we shall be brief in recalling them. The mass in the vector representation
in terms of the KK momenta and winding numbers $m_i,m^i$ ($i=1\ldots d$), reads
\begin{subequations}
\begin{eqnarray}
\label{vectormass}
\mathcal{M}^2 (\vect) &=& m\cdot e^t e \cdot  m = \tilde m_i
g^{ij} \tilde m_j + m^i g_{ij} m^j \ ,\qquad\\
\label {veccons} k&=& m_i m^i =0\ ,
\end{eqnarray}
\end{subequations}
where the last equation records the (quadratic) half-BPS condition
$m\wedge m=k=0$. Integer shifts of $B\to B+b$ induce a spectral
flow $m_i\to m_i - b_{ij} n^j$ on the lattice of BPS states,
leaving the dressed charge $\tilde m_i=m_i + B_{ij} m^j$ invariant
and preserving the condition $m\wedge m=0$. For the spinor
representation with $2^{d-1}$ charges
$(m^{[1]},m^{[3]},m^{[5]},\ldots)$\footnote{Integer subscripts or
superscripts in square brakets denote the number of antisymmetric
$SL(d)$ indices. When separated by a semi-colon as in
\eqref{dsconstr}, they stand for groups of antisymmetric indices
with no mutual symmetry property. The upper or lower position of
the indices denotes a gradient or contragradient representation of
$SL(d)$.} describing the wrapping numbers along the odd cycles of
$T^d$, the charges can be encapsulated in a differential form
$m=m^i dx_i + \frac{1}{3!}m^{ijk} dx_i\wedge dx_j\wedge dx_k +
\ldots $ and the effect of the $B$-field is to boost the charges
as $\tilde m = \exp\left(\frac{1}{2}B_{ij}dx^i \wedge dx^j \cdot
\right) m$, where $\cdot$ denotes the inner product. The formula
\cite{Pioline:1997pu}
\begin{subequations}
\label{dstring}
\begin{eqnarray}
\M^2 (\spi ) &=& \frac{1}{V_d}\left[ (\tilde m^i)^2 +
\frac{1}{3!} (\tilde m^{ijk})^2 +
\frac{1}{5!}(\tilde m^{ijklm})^2 +\dots \right] \\
\tilde m^i &=& m^i +  \frac{1}{2} m^{jki} B_{jk} + \frac{1}{8} m^{jklmi}
B_{jk} B_{lm}+\dots \\
\tilde m^{ijk} &=& m^{ijk} +  \frac{1}{2} m^{lmijk} B_{lm} + \dots\\
\tilde m^{ijklm} &=& m^{ijklm} + \dots
\end{eqnarray}
\end{subequations}
gives, up to a power $V_d/(g_s^2 l_s^8)=l_P^{d-8}$ of the T-duality invariant
Planck length and subject to the half-BPS conditions
\begin{subequations}
\label{dsconstr}
\begin{eqnarray}
k^{[4]}&=&k^{ijkl}=m^{[i}m^{jkl]}=0 \label{13con} \\
k^{[1;5]}&=&k^{i;jklmn}=m^{i[jk}m^{lmn]} +m^{i[jklm} m^{n]}=0\\
k^{[2;6]}&=&k^{ij;klmnpq}=m^{ij[k} m^{lmnpq]}=0\ ,
\end{eqnarray}
\end{subequations}
the mass of type IIB D-branes wrapped on an odd-dimensional cycle,
or the tension of type IIA D-branes wrapped on an odd-dimensional
cycle. Here we made explicit the constraints up to $d=6$ only. In
particular, we note that the first occurrence of the quadratic
constraints is for $d=4$, in which case they reduce to a singlet.
For $d=5$ they form a vector $\irrep{5}$, while for $d=6$ they
transform in an antisymmetric representation $\irrep{66}$ of the
T-duality group $SO(6,6,\Zint)$. More generally, one should
require the representation $\R\otimes_s\R$ to be irreducible.
Similarly, for the conjugate spinor representation with wrapping
numbers $m=(m,m^{[2]},m^{[4]},\ldots)$ around the even cycles of
$T^d$, we have
\begin{subequations}
\label{dparticle}
\begin{eqnarray}
\M^2 (\spb) &=& \frac{1}{V_d} \left[ \tilde m^2 +
\frac{1}{2} (\tilde m^{ij})^2 +
\frac{1}{4!}(\tilde m^{ijkl})^2 +\dots \right] \\
\tilde m &=& m +  \frac{1}{2} m^{ij} B_{ij} + \frac{1}{8} m^{ijkl}
B_{ij} B_{kl}+\dots \\
\tilde m^{ij} &=& m^{ij} +  \frac{1}{2} m^{klij} B_{kl} + \dots\\
\tilde m^{ijkl} &=& m^{ijkl} + \dots
\end{eqnarray}
\end{subequations}
with half-BPS conditions
\begin{subequations}
\label{dpconstr}
\begin{eqnarray}
k^{ijkl} & = & m^{[ij} m^{kl]} + m~m^{ijkl} = 0 \label{22con}
\\ k^{i;jklmn} & = & m^{i[j} m^{klmn]} + m~m^{ijklmn} =  0\\
k^{ij;klmnpq} & = &n^{ij} n^{klmnpq} + n^{ij[kl}n^{mnpq]}=  0
\end{eqnarray}
\end{subequations}
This describes the tension of type IIB D-branes wrapped on
an even-dimensional cycle, or the mass of type IIA D-branes
wrapped on an even-dimensional cycle.

With these T-duality invariant building blocks in hand, we may now
define the Eisenstein series for each of these three representations
as
\begin{equation}
\eis{SO(d,d,\Zint)}{\R}{s} = \hat{\sum_{m}} \delta (m \wedge m)
[ \M^2 (\R) ]^{-s}
\end{equation}
where have used the labels $\R = \vect, \spi, \spb $ for the
vector, spinor and conjugate spinor representations. Here $\delta
(m \wedge m)$ stands for the quadratic constraints
\eqref{veccons}, \eqref{dsconstr}, \eqref{dpconstr}, and $\M^2
(\R)$ are the mass formulae given in \eqref{vectormass},
\eqref{dstring}, \eqref{dparticle}. Not surprisingly, an explicit
computation (see Appendix \ref{eig})  shows that these Eisenstein
series are indeed eigenmodes of the Laplacian on the scalar
manifold \eqref{som},
\begin{subequations}
\begin{equation}
\Delta_{SO(d,d)} \eis{SO(d,d,\Zint)}{\R}{s}=
\Delta (\R,s) \eis{SO(d,d,\Zint)}{\R}{s}
\end{equation}
\begin{equation}
\label{solap}
\Delta_{SO(d,d)}=
\frac{1}{4} 
g_{ik} g_{jl} \left[ \frac{\partial}{\partial g_{ij}}
\frac{\partial}{\partial g_{kl}}
+ \frac{\partial}{\partial B_{ij}} \frac{\partial}{\partial B_{kl}}
\right]
+ \frac{1}{2}
g_{ij} \frac{\partial}{\partial g_{ij}}
\end{equation}
\end{subequations}
where the eigenvalues are given by
\begin{equation}
\label{soeigval}
\Delta (\vect,s) = s (s-d +1) \ ,\qquad
\Delta (\spi,s) = \Delta (\spb,s) = \frac{s d (s-d +1)}{4}\ ,
\end{equation}
in agreement with Eq. \eqref{eisgen}. The degeneracy of the spinor
and conjugate spinor (as well as the vector for $d=4$) is a
consequence of the outer automorphism which relates the two (or
the three for $d=4$, due to triality). We emphasize that the
derivation shows that the quadratic 1/2 BPS constraints are
essential for these Eisenstein series to be eigenmodes. For
instance, in the case of the vector representation, the analogue
of \eqref{delexp} is
\begin{equation}
 \Delta_{SO(d,d)} e ^{-m^tM(\vect) m /t}
= \left[ \frac{(m^tM(\vect) m)^2-4(m\wedge m)^2}{t^2}
- d \frac{m^tM(\vect) m}{t} \right] e ^{-m^tM(\vect) m /t}
\end{equation}
where $m\wedge m=m_i m^i$ vanishes on half-BPS states only.
For low dimensional cases however, the constraints drop or can be
solved, so that we are back to ordinary Eisenstein series.
This includes the $d=1$ vector series,
\begin{equation}
\eis{SO(1,1,\Zint)}{\vect}{s}=2\zeta(2s) \left( R^{2s} + R^{-2s} \right)
\end{equation}
or the $d<4$ spinor series,
\begin{subequations}
\label{eisso}
\begin{gather}
\label{eisso11} \eis{SO(1,1,\Zint)}{\spi}{s}=2\zeta(2s) R^{-s} \ ,\qquad
\eis{SO(1,1,\Zint)}{\spb}{s}=2\zeta(2s) R^{s} \\
\eis{SO(2,2,\Zint)}{\spi}{s}
=\eis{SL(2,\Zint)}{2}{s}(U) \ ,\qquad
\label{eisso22}
\eis{SO(2,2,\Zint)}{\spb}{s}
=\eis{SL(2,\Zint)}{2}{s}(T)\\
\label{eisso33}
\eis{SO(3,3,\Zint)}{\spi}{s}=
\eis{SL(4,\Zint)}{4}{s}  \ ,\qquad
\eis{SO(3,3,\Zint)}{\spb}{s}=
\eis{SL(4,\Zint)}{\bar 4}{s}
\end{gather}
\end{subequations}
where the identities in the last two lines follow from the
local isomorphisms $SO(2,2,\Real)=SL(2,\Real)\times SL(2,\Real)$
($U$ and $T$ denote
the standard complex moduli $U = (g_{12} + i V_2 )/g_{11}$, $T= B_{12} +
i V_2$ ) and $SO(3,3,\Real)=SL(4,\Real)$.

\subsection{One-loop modular integral and method of orbits}

Under toroidal compactification on a torus $T^d$, any string
theory exhibits the T-duality symmetry $SO(d,d,\Zint)$, and all
amplitudes should be expressible in terms of modular forms of this
group. For half-BPS saturated couplings, the one-loop amplitude
often reduces to an integral of a lattice partition function over
the fundamental domain $\F$ of the moduli space of genus-1 Riemann
surfaces,\footnote{This integral is divergent in the infrared region $\tau_2\to\infty$ when $d\geq 2$.
It may be regulated in many different ways, see e.g. \cite{Dixon:1991pc} for $d=2$. Different regulation schemes lead to results differing by an additive, moduli independent constant, which we ignore in our analysis.}
\begin{equation}
\label{oneloop} I_d = 2\pi \int_\F \frac{d^2\tau}{\tau_2^2} Z_{d,d}
(g,B;\tau)
\end{equation}
where $Z_{d,d}$ is the partition function (or theta function)
of the even self-dual
lattice describing the toroidal compactification,
\begin{subequations}
\begin{equation}
Z_{d,d} = V_d \sum_{m^i,n^i} e^{ -\frac{\pi}{\tau_2} (m^i + \tau
n^i ) (g_{ij}+B_{ij}) (m^j + \bar \tau n^j) } = (\tau_2)^{d/2}
\sum_{m_i,n^i} e^{ -  \pi \tau_2 \M^2 (\vect) -2\pi i \tau_1
m\wedge m }
\end{equation}
\end{subequations}
This is for instance the case for $R^4$ couplings in type II
strings on $T^d$, or $R^2$ or $F^2$ couplings in type II on
$K_3\times T^2$. In the above formula, a Poisson resummation on
the integers $m^i$ takes from the Lagrangian representation,
manifestly invariant under the genus 1 modular group, to the
Hamiltonian representation, manifestly invariant under T-duality.

It is natural to expect a connection between this one-loop modular
integral and the $SO(d,d,\Zint)$ Eisenstein series defined above.
As is well known, the $\tau$-integral can be carried out by the
method of orbits, which corresponds to a large volume expansion of
the integral. This was first carried out in \cite{Dixon:1991pc}
and extended to higher dimensional tori in
\cite{Kiritsis:1997em,Kiritsis:1997hf}. We will briefly review
these results for later comparison with the Eisenstein series.

In order to carry out the integral on the fundamental domain of
the upper half plane, one uses the fact that an $SL(2,\Zint)$
modular transformation on $\tau$ can be reabsorbed by an
$SL(2,\Zint)$ action on the doublet $(m^i,n^i)$: one can thus
restrict the sum over $(m^i,n^i)$ to a sum over their
$SL(2,\Zint)$ orbits, while unfolding the integration  to a larger
domain depending on the centralizer of the orbit. The orbits can
be classified by defining the sub-determinants, $d^{ij}=\frac12(m^i
n^j-m^j n^i)$, so that $d^{ij}$ is a $d\times d$, rank 2 antisymmetric
matrix. We then have the {\it trivial orbit}, $m^i=n^i=0$, with a
contribution
\begin{equation}
I^{\rm tr}_{d}=2\pi V_d \int_{\F} \frac{d^2 \tau}{\tau_2^2}
=\frac{2\pi^2}{3}V_d\ ;
\label{triv}\end{equation}
the {\it degenerate} orbits, with all $d$'s being zero:
in this case we can set $n^i=0$, unfold the integration domain
$\F$ onto the strip $\tau_1\in[-\frac{1}{ 2},\frac{1}{ 2}],\tau_2\in
\Real^+$, and carry out the integrals:
\begin{equation}
I^{\rm d}_{d}=2V_d \hat{\sum_{m^i}}\frac{1}{ m^{i}
g_{ij}m^{j}}= 2V_d \eis{SL(d,\Zint)}{\irrep{d}}{s=1}(g_{ij})\ ;
\label{deg}\end{equation}
the {\it non-degenerate} orbits, where at least one of the $d^{ij}$
is non-zero. The $SL(2,\Zint)$ modular action can be completely fixed
in order to unfold the integration domain to twice the upper-half
plane.
After Gaussian integration on $\tau$, we obtain:
\begin{equation}
I^{\rm n.d}_{d}=4\pi V_d \bar{\sum_{m^i,n^i}}
\frac{\exp\left[-2\pi \sqrt{(m\cdot g \cdot m)(n\cdot g\cdot
n)-(m\cdot g\cdot n)^2}+2\pi i  B_{ij} m^i n^j\right]
}{\sqrt{(m\cdot g\cdot m)(n\cdot g\cdot n)-(m\cdot g\cdot n)^2}}
\label{nondeg}\end{equation} The summation is performed over all
sets of $2n$ integers, having at least one non-zero $d^{ij}$,
modded out by the $SL(2,\Zint)$ modular action (for $d=1$, this is
$m>0,0\leq n<m$). These terms are all exponentially suppressed at
large $V_d$, albeit not in a uniform way.

For low-dimensional cases, the sum can be further simplified, and
yields the well-known results:
\begin{equation}
I_1 = \frac{2 \pi^2}{3} \left( R+ \frac{1}{R} \right)\ ,\qquad
I_2 =-2\pi \log ( T_2 U_2|\eta(T)\eta(U)|^4 )
\end{equation}
It is remarkable that these results can be rewritten in terms
of $SO(d,d,\Zint)$ Eisenstein series. Indeed, using the properties
\eqref{eisso} and \eqref{ded}, we find
\begin{prop}
For $d=1,2$, the one-loop integral $I_d$ in \eqref{oneloop}
can be rewritten as the sum of the $SO(d,d,\Zint)$ Eisenstein series
of order 1 in the spinor and conjugate spinor representations:
\begin{equation}
\label{cc1}
I_d = 2 \eis{SO(d,d,\Zint)}{\spi}{s=1} + 2 \eis{SO(d,d,\Zint)}{\spb}{s=1}
\end{equation}
\end{prop}
In particularly, the result is manifestly invariant under the extended
T-duality $O(d,d,\Zint)$, where the extra generator exchanges the
two spinors. We shall now substantiate a similar claim
for $d>2$, by showing that the two sides are eigenmodes of second
order differential operators with the same eigenvalues, and
that they also agree in various limits. At this point, we note
that the fact that the two spinor representations contribute
is in agreement with the invariance of the modular integral
under the extended group $O(d,d,\Zint)$ which exchanges the
two spinors. Besides, for $d=1$ the vector Eisenstein series
$-\frac{\pi ^2}{3}\eis{SO(1,1,\Zint)}{\vect}{s=-1/2}$ is
an equally valid candidate.

\subsection{A new second order differential operator}

Given that our Eisenstein series are eigenmodes of the
$SO(d,d)$ Laplacian \eqref{solap}, we should ask
about its action on the modular integral \eqref{oneloop}.
An explicit computation of the action of $\Delta_{SO(d,d)}$
on the integrand shows that
the lattice sum satisfies the differential equation
\begin{equation}
\label{soneq} \left[ \Delta_{SO(d,d)}-2\Delta_{SL(2)} +
\frac{d(d-2)}{4} \right] Z_{d,d} = 0
\end{equation}
where $\Delta_{SL(2)}=\frac{1}{2}\tau_2^2 \left(
\frac{\pa ^2}{\pa \tau_2^2}+\frac{\pa ^2}{\pa \tau_1^2} \right)$
is the Laplacian on the upper-half plane. Upon integrating by
parts the second term, we get a boundary term which vanishes, so that
the modular integral itself is an eigenmode of $\Delta_{SO(d,d)}$:
\begin{equation}
\Delta_{SO(d,d)}  I_d = \frac{d(2-d)}{4} I_d
\end{equation}
The modular integral $I_d$ is therefore degenerate with the
$SO(d,d,\Zint)$ Eisenstein series
\begin{equation}
\label{degsoo1}
\eis{SO(d,d,\Zint)}{\vect}{s=d/2-1}\ ,\quad
\eis{SO(d,d,\Zint)}{\spi}{s=1}\ ,\quad
\eis{SO(d,d,\Zint)}{\spb}{s=1}\ ,\quad
\end{equation}
or their ``duals''
\begin{equation}
\label{degsoo2}
\eis{SO(d,d,\Zint)}{\vect}{s=d/2}\ ,\quad
\eis{SO(d,d,\Zint)}{\spi}{s=d-2}\ ,\quad
\eis{SO(d,d,\Zint)}{\spb}{s=d-2}\ ,\quad
\end{equation}
We expect the functions in \eqref{degsoo2}
to be related to the ones in \eqref{degsoo1}
by a duality transformation analogous to \eqref{slinv},
although we cannot prove this statement at present
due to the presence of constraints.

Less expected however is the existence of a second differential
operator  $\square_d$, involving only the metric, which also
annihilates the integrand up to a total derivative:
\begin{equation}
\label{box} \square_d= \Delta_{Gl(d)} - \frac{1}{8} \left(
g_{ij} \frac{\pa}{ \pa g_{ij} } \right)^2
=
\Delta_{SL(d)} + \frac{2-d}{8d} \left( 
g_{ij} \frac{ \pa}{\pa g_{ij} } \right)^2
\end{equation}
where the $Gl(d)$ and $SL(d)$ Laplacians are given in \eqref{gllap},
\eqref{sllap}.
Indeed an explicit computation shows that
\begin{equation}
\label{sqeq}
\left[ \square_d - \Delta_{SL(2)}+ \frac{d(d-2)}{8} \right] Z_{d,d} =
0\ ,\qquad \mbox{\ie}\quad  \square_d I_d = \frac{d(2-d)}{8} I_d.
\end{equation}
The operator $\square_d$ is non-invariant under $SO(d,d)$, but is
invariant under complete inversion of the metric. \footnote{In
Ref.\cite{Kiritsis:1995iu} it was shown that the one-loop integral
is an eigenfunction under a non-invariant second order operator
$\Delta$ that involves both $g$ and $B$. The relation with
$\square_d$ in \eqref{box} is $\square_d =
\Delta_{SO}-\frac{1}{2}\Delta+\frac{d(d-2)}{8}$. The equation
\eqref{soneq} involving $\Delta_{SO}$ was also given in Ref.
\cite{Kiritsis:1997hf}.}

This last property gives a strong constraint for the identification
of the modular integral $I_d$ with Eisenstein series. Indeed, one
can show that the spinor Eisenstein series are eigenmodes of
$\square_d$ for $s=1$ only, whereas the vector is always an
eigenmode:
\begin{subequations}
\label{sqeigval}
\begin{gather}
\square_d \eis{SO(d,d,\Zint)}{\vect}{s} =
 \frac{s(s-d+1)}{2} \eis{SO(d,d,\Zint)}{\vect}{s} \\
\square_d \eis{SO(d,d,\Zint)}{\spi}{s=1} = \frac{d(2-d)}{8}
\eis{SO(d,d,\Zint)}{\spi}{s=1} \label{bspieig}  \\ \square_d
\eis{SO(d,d,\Zint)}{\spb}{s=1} =
\frac{d(2-d)}{8}\eis{SO(d,d,\Zint)}{\spb}{s=1} \label{bspbeig}
\end{gather}
\end{subequations}
In particular, we see that the spinor Eisenstein series of order $s=1$
and $s=d-2$ are distinct, even though they are degenerate under
$\Delta_{SO(d,d)}$. A peculiarity occurs for $d=4$, where
the spinor Eisenstein series is an eigenmode for all $s$,
whereas the conjugate spinor is an eigenmode for $s=1$ only:
\begin{equation}
\label{so44}
\square_4(\vect,s)=\square_4(\spi,s)=\frac{s(s-3)}{2}\ ,\qquad
\square_4(\spb,s=1)=-1
\end{equation}
The three $SO(4,4,\Zint)$ Eisenstein series at $s=1$
are therefore degenerate under both $\Delta_{SO}$ and $\square_4$,
and we conjecture that they are actually equated by triality.
The degeneracy is however lifted at $s\ne 1$.

Summarizing the results in this section, we see that the only
candidates for representing the modular integral \eqref{oneloop}
are the order $s=1$ spinor and conjugate spinor series, together with
the order $s=d/2-1$ vector series and their duals. In order to sort out
these possibilities, we need to determine the behaviour of these
invariant functions in various limits.

\subsection{Large volume behaviour}
The large volume limit of the modular integral $I_d$ has already
been obtained from the orbit decomposition. The behaviour of the
Eisenstein series on the other hand can be obtained by Poisson
resummation techniques similar to the ones described in Section 2,
with the complication of the constraints. The actual computation
is deferred to Appendix \ref{lve}, and we present only the
results, specializing to the relevant value of $s$. In the case of
the vector representation, we are able to determine the complete
large volume expansion:
\begin{multline}
 \eis{SO(d,d,\Zint)}{\vect}{s=\frac{d}{2}-1}=
\frac{\pi^{\frac{d}{2}-2}}{\Gamma(\frac{d}{2}-1)} \left( V_d
\eis{SL(d,\Zint)}{d}{s=1}(g_{ij}) + \frac{\pi^2}{3} V_d \right.  +
\\ \left. + 2\pi V_d \bar{\sum_{m^i,n^i} } \frac{ \exp\left(-2\pi
\sqrt{|(m\cdot g\cdot m)(n\cdot g \cdot n)-(m\cdot g\cdot n)^2|} +
2\pi i B_{ij} m^i n^j \right) } {\sqrt{|(m\cdot g\cdot m)(n\cdot g
\cdot n)-(m\cdot g \cdot n)^2}|} \right)
\label{largevolV}\end{multline} Here the sum runs over
non-degenerate $SL(2,\Zint)$ orbits of $(m^i,n^i)$. Comparing with
the expansion $I_d^{\rm tr}+I_d^{\rm d}+I_d^{\rm n.d.}$ of the
modular integral \eqref{oneloop} in Equations
\eqref{triv}-\eqref{nondeg}, we see a complete matching and thus
obtain the theorem
\begin{theo}
The integral $I_d$ \eqref{oneloop} of the $(d,d)$ lattice
partition function on the fundamental domain of the moduli space of
genus 1 Riemann surfaces is given for $d\ge 3$ by the $SO(d,d,\Zint)$
Eisenstein series of order $s=d/2-1$ in the vector representation
\begin{equation}
\label{c2}
I_d = 2 \frac{\Gamma(\frac{d}{2}-1)}{\pi^{\frac{d}{2}-2}}
\eis{SO(d,d,\Zint)}{\vect}{s=\frac{d}{2}-1}
\end{equation}
\end{theo}
This provides a convenient representation of the one-loop integral
$I_d$, manifestly invariant under T-duality. The Eisenstein series
of order $s=d/2$ in the vector representation is degenerate with
the one above, but singular, so we ignore it here.

In the case of the Eisenstein series of order 1 in the spinor
representations, the determination of the asymptotic behaviour
is complicated by the presence of the constraints, and we
have to content ourselves with the partial results
\begin{subequations}\label{largevol}
\begin{eqnarray}
\eis{SO(d,d,\Zint)}{\spi}{s=1} &=&  V_d
\eis{SL(d,\Zint)}{d}{s=1}(g_{ij}) + \frac{\pi^2}{3}V_d + \dots
\label{largevolS} \\
\eis{SO(d,d,\Zint)}{\spb}{s=1} &=& \frac{\pi
^2}{3} V_d +
\pi V_d \eis{SL(d,\Zint)}{[2]}{s=1/2}(g_{ij})
+ \dots
\label{largevolC}
\end{eqnarray}
\end{subequations}
to be compared with
\begin{equation} I_d = \frac{2\pi^2}{3} V_d + 2 V_d
\eis{SL(d,\Zint)}{d}{s=1}(g_{ij}) + \dots \label{largevolI}
\end{equation}
The second term in \eqref{largevolS} is correct
for $d\leq 3$, but we are not able to prove  it explicitly for
$d>3$, due to the presence of the constraints; there
are also exponentially suppressed corrections that we
did not write. The second term in \eqref{largevolC}
denotes the Eisenstein series
of $SL(d,\Zint)$ in the antisymmetric representation, and appears
only when $d\ge 2$. For the particular order $s=1/2$, it is easy to
check from \eqref{casanti} that this series has the same eigenvalue
as the Eisenstein series of order 1 in the fundamental
representation. For $d=2,3$ we have also explicitly checked the equality
of the two Eisenstein series, so that we are led to assert
\begin{guess}
For any $d$, the Eisenstein series of $SL(d,\Zint)$ in the antisymmetric
2-tensor representation $\irrep{[2]}$ at the particular order $s=1/2$ coincides with
the Eisenstein series of order 1 in the fundamental
representation:
\begin{equation}
\label{c0}
\eis{SL(d,\Zint)}{\irrep{[2]}}{s=1/2}
= \frac{1}{\pi}
\eis{SL(d,\Zint)}{\irrep{d}}{s=1}
\end{equation}
\end{guess}
Assuming this is true, we can now formulate
our second claim for the one-loop threshold:
\begin{guess} The integral \eqref{oneloop} of the $(d,d)$ lattice
partition function on the fundamental domain of the moduli space of
genus 1 Riemann surfaces is given for $d\ge 3$ by the $SO(d,d,\Zint)$
Eisenstein series of order $s=1$ in any of the two spinor representations:
\begin{equation}
\label{c1}
I_d = 2 \eis{SO(d,d,\Zint)}{\spi}{s=1} = 2 \eis{SO(d,d,\Zint)}{\spb}{s=1}
\end{equation}
\end{guess}
This is to be contrasted with the $d=1,2$ case \eqref{cc1},
where the two spinors contribute in order to enforce the
$O(d,d,\Zint)$ invariance of the integral \eqref{oneloop}. When $d>2$,
we conjecture that the two Eisenstein series are equal for the
particular order $s=1$, so that a single series is sufficient
to reproduce the threshold. For $d=3$, this
conjecture is actually a theorem, as follows from the computation
of $R^4$ couplings in 7 dimensions \cite{Kiritsis:1997em}.
For $d=4$, the conjecture \eqref{c1} together with the
theorem \eqref{c2} implies that the one-loop integral \eqref{oneloop}
is invariant under $SO(4,4)$ triality, a fact not obvious from
its representation as a theta function.

\subsection{Asymmetric thresholds and elliptic genus}
So far, we focused on symmetric thresholds of the type
\eqref{oneloop}, which often appear for half-BPS saturated
couplings in type II strings, and showed how they could be
expressed in terms of Eisenstein series of the $SO(d,d,\Zint)$
duality group. For heterotic strings however, the BPS condition
constrains only the left-movers to be in their ground states, and
the amplitude usually involves all excitations of the right-moving
oscillators. Here we want to investigate the possible relevance of
Eisenstein series for these quantities. Even though the
--negative-- outcome can already be anticipated due to the issue
of symmetry enhancement, this will allow us to establish some
identities that may become useful in later studies.

One-loop BPS-saturated couplings for toroidal compactifications of
the heterotic string can usually be written as
the modular integral
\begin{equation}
{\cal I}^{\rm het}  =
  \int_{ {\cal F}}\frac{d^2\tau}{\tau_2^2}
\; Z_{d,d} (g,B;\tau) \ {\cal A}(F,R, \tau)
\end{equation}
where the insertion ${\cal A}$ is an almost holomorphic modular
form of weight 0, depending on the background gauge-field $F$
and curvature $R$ in the uncompact dimensions. By almost
holomorphic, we mean that ${\cal A}$ can be expanded as
a finite polynomial in $1/\tau_2$
\begin{equation}
{\cal A}(F,R,\tau) =  \sum_{\nu=0}^{\nu_{\rm max}}\
\frac{1}{\tau_2^\nu}
{\cal A}^{(\nu)}(F,R,\tau)
\label{exp}
\end{equation}
with $ {\cal A}^{(\nu)}(F,R,\tau)$ a meromorphic function in
$q = e^{2\pi i \tau}$.
The non-holomorphic contributions $\nu\ge 1$ come from back-reaction
effects, or equivalently from contact terms at the boundary of
moduli space. In all string applications, the coefficients
$ {\cal A}^{(\nu)}$ have
Laurent expansions with at most a simple pole in $q$,
arising from the left-moving tachyon.

When the elliptic genus does not depend on the gauge fields,
it is actually possible to switch on Wilson lines $Y$, giving
an $SO(d,d+k,\Zint)$-invariant threshold
\begin{equation}
\label{oneloopk}
I_{d,k} = \int_{ {\cal F}}\frac{d^2\tau}{\tau_2^2}\;
  Z_{d,d+k} (g,B,Y;\tau) \
{\cal A}_k(R, \tau)
\end{equation}
where
\begin{subequations}
\begin{equation}
  Z_{d,d+k}
= (\tau_2)^{d/2}  \sum_{m_i,p_I,n^i} e^{ -  \pi \tau_2 \M^2
(\vect) - 2\pi i \tau_1 v^t \eta v   }
\end{equation}
\begin{equation}
\M^2 (\vect) = v^t M_{d,k} (\vect) v \ ,\quad v =(m_i,p_I,n^i)\
,\quad i
=1\ldots d\ ,\quad I = 1 \ldots k \ 
\end{equation}
\end{subequations}
and ${\cal A}_k$ is now an almost holomorphic modular form of
weight $-k/2$. We can derive, also in this case, a set of second
order partial differential equations satisfied by the lattice
partition function $Z_{d,d+k}$. It is convenient to choose the
following Iwasawa gauge, in the basis where the $SO(d,d+k)$
invariant tensor reads $\eta=\scriptstyle\begin{pmatrix}
&&1_d\\&1_k&\\ 1_d&&\end{pmatrix}$:
\begin{equation}
M_{d,k} (\vect) =
\begin{pmatrix} 1 & & \\ Y^t & 1 & \\ C^t & -Y & 1  \end{pmatrix} \cdot
\begin{pmatrix}g^{-1} & & \\ & 1_k & \\ & & g     \end{pmatrix} \cdot
\begin{pmatrix} 1 & Y & C \\ & 1 & -Y^t \\ & & 1 \end{pmatrix}
\end{equation}
with $C=B-YY^t/2$ and $B$ antisymmetric,
as a result of the $SO(d,d+k)$ constraint
$M_{d,k}^t\eta M_{d,k}=\eta$.
The right action by the $SO(d,d+k)$ elements
\begin{equation}
\begin{pmatrix} 1 & y & -yy^t/2 \\ & 1 & -y^t \\ & & 1
\end{pmatrix} \; \; \mbox{and } \; \;
\begin{pmatrix} 1 &   & b \\ & 1 &  \\ & & 1 \end{pmatrix}
\end{equation}
preserving the Iwasawa gauge
generates a set of continuous Borel symmetries
\begin{equation}
Y\to Y+y\ ,\quad B\to B+ \frac{1}{2} ( y Y^t - Y y^t)\qquad \mbox{ or }
B\to B+b
\end{equation}
which reduces to a discrete subgroup at the quantum level. The
Laplacian then takes the form
\begin{equation}
\label{asymlap}
\Delta_{SO(d,d+k)}= \frac{1}{4} 
g_{ik} g_{jl}
\left[ \frac{\partial}{\partial g_{ij}} \frac{\partial}{\partial g_{kl}}
+  \frac{\pa}{ \pa B_{ij}} \frac{\pa}{  \pa B_{kl}}
\right] + \frac{2-k}{4} 
g_{ij} \frac{\partial}{\partial g_{ij}}
+\Delta_Y
\end{equation}
where
\begin{equation}
\Delta_Y=\frac{1}{2} g_{ij} \delta^{IJ}
\left[ \frac{\pa}{ \pa Y_i{}^I} - \frac{1}{2} Y_k^I \frac{\pa}{ \pa
    B_{ik}} \right]
\left[ \frac{\pa}{ \pa Y_j{}^J} - \frac{1}{2} Y_l^J \frac{\pa}{ \pa
    B_{jl}} \right]
\end{equation}
We may then show that the lattice partition function satisfies
the following identities, generalizing \eqref{soneq} and \eqref{sqeq},
\begin{subequations}
\label{sonkeq}
\begin{equation}
\left[\Delta_{SO(d,d+k)} + \frac{d(d+k -2 )}{4}  - {\cal{D}} \right]
Z_{d,d+k} =0
\end{equation}
\begin{equation}
\left[\square_{d,k} + \frac{d(d+k -2 )}{8} - \frac{1}{2} {\cal{D}}
\right] Z_{d,d+k} =0
\end{equation}
\end{subequations}
where $\cal D$ is the modular-covariant second order differential operator
acting on modular forms of weight $(k/2,0)$, and
$\square_{d,k}$ generalizes the non-invariant operator \eqref{box}:
\begin{subequations}
\begin{equation}
{\cal{D}} = 4 \tau_2^2 \pa_{\bar{\tau}} D_\tau \ ,\qquad
D_\tau = \pa_\tau -  i \frac{k}{4\tau_2 }
\end{equation}
\begin{equation}
\square_{d,k}=\frac{1}{4}  
g_{ik} g_{jl}
\frac{\partial}{\partial g_{ij}} \frac{\partial}{\partial g_{kl}}
-\frac{1}{8}\left( 
g_{ij} \frac{\partial}{\partial
g_{ij}} \right)^2 +\frac{d+1-k/2}{4} 
g_{ij}\frac{\partial}{\partial g_{ij}} +\frac{1}{2}\Delta_Y
\end{equation}
\end{subequations}

\subsection{Boundary term and symmetry enhancement}

A quick glance at the partial differential equations
\eqref{sonkeq} may lead us to the conclusion that the
$SO(d,d+k)$-invariant one-loop integral \eqref{oneloopk} should
again be an eigenmode of the operators $\Delta_{SO(d,d+k)}$ and
$\square_{d,k}$. This is wrong however, due to the presence in
${\cal A}$ of the tachyonic pole in $1/q$. This pole is usually
killed by the integration on the strip $\tau_1\in[-1,1]$ (at large
$\tau_2$), except at special points of the moduli space where the
lattice contains a length 2 vector: the contribution $q^{1}\bar
q^0$ from the lattice sum cancels the pole, which signals an
enhancement of the gauge symmetry in space-time. In particular,
using the  identity \eqref{sonkeq} requires particular care for
the boundary term
\begin{equation}
-\int_{\cal F} d^2\tau \frac{\pa}{\pa \bar\tau} \left[
\frac{1}{\tau_2^\nu}\
{\cal A}_k^{(\nu)}(F,R)
\frac{\pa}{\pa \tau}( Z_{d,d+k})\
\right]
=
\lim_{\tau_2\to\infty}
(\tau_2)^{d/2-\nu-1}
\left[ {\cal A}_k^{(\nu)}(F,R)
\frac{Z_{d,d+k}}{(\tau_2)^{d/2}} \right]_{q^0\bar q^0}
\end{equation}
The contribution from the constant term in ${\cal A}$ and the
ground state of $Z_{d,d+k}/(\tau_2)^{d/2}$ yields a
moduli-independent divergent (for $d/2-\nu-1>0$) term which
implies a harmless non-harmonicity, whereas the pole term in
${\cal A}$ generates a harmonic anomaly localized at enhanced
symmetry points in the moduli space, clearly not captured by any
candidate Eisenstein series. Finally, the term integrated by parts
now involves the descendant of the elliptic genus, as explained in
\cite{Kiritsis:1997hf}. For $d=2$, the answer to this problem is
well known: the threshold involves the automorphic form of
$SO(2,2+k)$ constructed by Borcherds
\cite{Borcherds:1995,Borcherds:1996} (see also
\cite{Harvey:1996fq,Kiritsis:1997hf} in the physics literature).
The evaluation of the modular integral \eqref{oneloopk} by the
method of orbits gives a presentation of this form as an infinite
product over a sublattice, and each term vanishes on a particular
divisor of $[SO(2)\times SO(k)]\backslash SO(2,2+k)$ where the
gauge symmetry is enhanced. It would be interesting to construct
the generalization of these objects to $d>2$, where the complex
structure is not present (and find the analogue of the generalized
prepotentials obtained in Ref.\cite{Kiritsis:1997hf}) but we will
not pursue this line here.

\section{U-duality and non-perturbative $R^4$ thresholds \label{npert} }
While Eisenstein series provide a nice way to rewrite one-loop
integrals such as \eqref{oneloop}, their utility becomes even more
apparent when trying to extend the perturbative computation into
a non-perturbatively exact result. Indeed, a prospective exact
threshold should reduce in a weak coupling expansion
to a sum of T-duality invariant Eisenstein-like perturbative terms, plus
exponentially suppressed contributions, and Eisenstein series of the
larger non-perturbative duality symmetry are natural candidates
in that respect. This approach was taken in \cite{Kiritsis:1997em} for $R^4$
couplings in type II theories toroidally compactified to $7,8,9$
dimensions, where the technology of $SO(d,d,\Zint)$ Eisenstein
series was hardly needed; here we would like to extend it
to lower dimensional compactifications, in an attempt to
understand non-perturbative effects in these cases as well.

\subsection{$R^4$ couplings and non-renormalization}
Four graviton $R^4$ couplings in maximally supersymmetric theories
have been argued in dimension $D\ge 8$ to receive no perturbative
corrections beyond the tree-level and one-loop terms, and we shall
assume that this holds in lower dimensions as well. The tree level
term is simply obtained by dimensional reduction of the
ten-dimensional $2\zeta(3)e^{-2\phi}$ term found in
\cite{Gross:1986iv}, while the one-loop term was explicitly shown
to be given by the modular integral \eqref{oneloop}, after
cancellation of the bosonic and fermionic oscillators, so that
\begin{equation}
\label{pertr4}
f_{R^4}= 2\zeta(3) \frac{V_d}{g_s^2} + I_d + \mbox{non pert.}
\end{equation}
While the Ramond scalars are decoupled from the perturbative
expansion by Peccei-Quinn symmetries, the full non-perturbative
result should depend on all the scalars in the symmetric space
$K\backslash \exc(\Real)$, where $\exc(\Real)$ is the maximally
non-compact real form (also known as the normal real form) of the
series of classical simply laced Lie groups $E_2=SL(2)$,
$E_3=SL(3)\times SL(2)$, $E_4=SL(5)$, $E_5=SO(5,5)$ and
exceptional Lie groups $E_6$, $E_7$, $E_8$
\cite{Cremmer:1979up,Julia:1997cy}. It should furthermore be
invariant under the discrete symmetry group $\exc(\Zint)$ also
known as the U-duality group \cite{Hull:1995ys}, which arises from
the T-duality $SO(d,d,\Zint)$ by adjoining the exchange of the
eleventh M-theory direction with any perturbative direction. The
moduli space $K\backslash \exc(\Real)$ has the structure of a
bundle on the manifold $[SO(d)\times SO(d)]\backslash
SO(d,d,\Real)$ on which the Neveu-Schwarz scalars $\phi,g,B$ live,
with a fibre transforming as a spinor representation of
$SO(d,d,\Real)$ in which the Ramond scalars live. For $D\ge 8$, it
was shown \cite{Pioline:1998mn,Green:1998by} that the exact
threshold is an eigenmode of the Laplacian on the full scalar
manifold as a consequence of supersymmetry, and we shall also
assume that this persists in lower dimensions.

As shown in conjecture 1, the one-loop contribution can
be written as the order $s=1$ $SO(d,d,\Zint)$ Eisenstein series
in the spinor representation. On the other hand, the tree-level term
can itself be represented as an Eisenstein series in the
singlet representation, using the property
\begin{equation}
\eis{G(\Zint)}{\irrep{1}}{s} = 2 \zeta (2s)
\end{equation}
valid for any $G$, which provides a natural representation for
Apery's transcendental number $\zeta(3)$. In analogy with the
$D\ge 8$ case, we do not expect any further perturbative
contribution. For $2<d<8$, the $R^4$ threshold should thus be an automorphic
form of $\exc(\Zint)$ with asymptotic behavior
\begin{equation}
f_{R^4}= \frac{V_d}{g_s^2} \eis{SO(d,d,\Zint)}{\irrep{1}}{s=3/2}
+ 2 \eis{SO(d,d,\Zint)}{\irrep{\spi}}{s=1} + \mbox{non pert.}
\end{equation}

\subsection{String multiplet and non-perturbative $R^4$ couplings}
In order to propose a non-perturbative extension of this result,
we therefore need to unify the singlet and spinor representations
of $SO(d,d,\Zint)$ into a representation of $\exc(\Zint)$.
Remarkably, there is one, namely the string multiplet, corresponding
to the leftmost node in the Dynkin diagram
\begin{equation}
\label{dynkin}
\begin{array}{ccccccccc}
&&&&\frac{1}{l_M^3}&&&&\\
&&&&\mid& &&&\\
\frac{R_1}{l_M^3}&-&
\frac{R_1 R_2}{l_M^6}&-&
\frac{R_1 R_2 R_3}{l_M^9}&-&
\frac{R_1 R_2 R_3 R_4}{l_M^9}&-\dots-&
\frac{1}{R_{d+1}}
\end{array}
\end{equation}
where each node is labelled by the tension of the states
transforming in the corresponding representation
\cite{Elitzur:1997zn,Obers:1998fb}. The string multiplet is
described by a collection of charges $m^{[1]},m^{[4]},m^{[1;6]}$
describing the wrappings of membranes, five-branes and
Kaluza--Klein mono\-poles respectively\footnote{For simplicity we
restrict ourselves to the case $d\leq 6$, \ie $D\geq 4$.}, with a
BPS mass given by
\begin{equation}
\label{tension}
\T^2 =   \frac{1}{l_M^6} \left( \tilde m^{[1]} \right)^2 +
         \frac{1}{l_M^{12}} \left( \tilde m^{[4]} \right)^2 +
         \frac{1}{l_M^{18}} \left( \tilde m^{[1;6]} \right)^2 \ .
\end{equation}
The dressed charges are given by
\begin{equation}
\label{cshft}
\begin{array}{lllllll}
\tilde m^{[1]}    &=& m^{[1]} &+& \C_3 m^{[4]} &+&
                 \left(  \C_3 \C_3 + \E_6 \right) m^{[1;6]}  \\
\tilde m^{[4]}    &=& m^{[4]} &+& \C_3 m^{[1;6]}
                &&  \\
\tilde m^{[1;6]}&=& m^{[1;6]} &&  && \\
\end{array}
\end{equation}
where $\C_3$ and $\C_6$ are the expectation value of the M-theory
three-form and its dual, to be supplemented with an extra
$\K_{1;8}$ form in $D=3$. See Ref. \cite{Dijkgraaf:1997cv} for the
$d \leq 4$ case and \cite{Obers:1997kk,Obers:1998fb} for the
general $d$ case. This amounts to an explicit partial Iwasawa
decomposition of the symmetric spaces $K\backslash\exc(\Real)$.
The corresponding state preserves half the supersymmetries
provided the following conditions are obeyed
\cite{Dijkgraaf:1997cv,Obers:1998fb}:
\begin{subequations}
\begin{eqnarray}
k^{[5]} &=& m^{1} m^{[4]} =0\label{so55cond}\\
k^{[2;6]} &=& m^{1} m^{[1;6]} + m^{[4]} m^{[4]}=0\\
k^{[5;6]} &=& m^{[4]} m^{[1;6]}=0
\end{eqnarray}
\end{subequations}
The above constraints in turn transform as a U-duality multiplet,
namely the three-brane multiplet \cite{Obers:1998fb}. For
completeness, Table \ref{stringm}  lists  the U-duality groups and
string multiplets for any $d \leq 6$.

\begin{table}
\begin{center}
\begin{tabular}{|c|c||c|c|l|l|}
\hline
$D$&$d+1$ & U-duality group & irrep & $SL(d+1)$ content &
$SO(d,d)$ content \\ \hline
10&1 & 1 & \irrep{1} & \irrep{1} & \irrep{1} \\
9 &2 & $SL(2,\Zint)$ & \irrep{2}  & \irrep{2} & \irrep{1} + \irrep{1}   \\
8 &3 & $SL(3,\Zint)\times SL(2,\Zint)$ & \irrep{(3,1)} & \irrep{3} &
\irrep{1} + \irrep{2} \\
7 &4 & $SL(5,\Zint)$ & \irrep{5} & \irrep{4} + \irrep{1} & \irrep{1}
+ \irrep{4} \\
6 &5 & $SO(5,5,\Zint)$ & \irrep{10} & \irrep{5} + \irrep{\bar 5} & \irrep{1}
+ \irrep{8_S} + \irrep{1}  \\
5 &6 & $E_{6(6)}(\Zint)$ & \irrep{\bar{27}} &
                   \irrep{6} + \irrep{15} + \irrep{\bar{6}} & \irrep{1} +
\irrep{16} + \irrep{10}    \\
4 &7 & $E_{7(7)}(\Zint)$ & \irrep{133} & \irrep{7} + \irrep{35} +
\irrep{28}+\dots
 &  \irrep{1} + \irrep{32} + (\irrep{1}+\irrep{66})+\dots\\
\hline
\end{tabular}
\end{center}
\caption{String multiplets of $\exc(\Zint)$.
\label{stringm}}
\end{table}

The decomposition of this $\exc (\Zint)$ irreducible
representation into $SO(d,d,\Zint)$ representations was carried
out in \cite{Obers:1997kk,Obers:1998fb}, and indeed gives a
singlet $m=m^s$, a spinor $\spi=(m^i,m^{sijk},m^{s,sijklm})$, plus
some other multiplets $\irrep{O}$ when $d\ge 4$. In particular,
for $d=4$, there is an extra singlet $\irrep{O}=m^{ijkl}$ of
$SO(4,4,\Zint)$, and for $d=5$ a vector
$\irrep{O}=(m^{ijkl},m^{i;jklmn})$ of $SO(5,5,\Zint)$. The mass
formula \eqref{tension} is easily rewritten, for vanishing RR
backgrounds, in terms of T-duality quantities, using the relations
$l_M^3=g_s l_s^3$, $R_s=g_s l_s$:
\begin{equation}
\T^2= m^2 + \frac{V_d}{g_s^2} \M^2(\spi) + \frac{V_d^2}{g_s^4}
\M^2(\irrep{O})
\end{equation}
where we set $l_s=1$ and $\M^2(\irrep{O})$ is the usual T-duality
invariant mass for a singlet ($d=4$) or a vector ($d=5$). Given
this group theory fact, it is therefore quite tempting to consider
the following non-perturbative generalization of \eqref{pertr4}:
\begin{guess}
The exact four-graviton $R^4$ coupling in toroidal compactifications
of type II theory on $T^d$, or equivalently M-theory on $T^{d+1}$,
is given, up to a factor of Newton's constant,
by the Eisenstein series of the U-duality group
$\exc(\Zint)$ in the {\it string multiplet} representation,
with order $s= 3/2$:
\begin{equation}
\label{cr4}
f_{R^4} = \frac{V_{d+1}}{l_M^9}
\eis{\exc(\Zint)}{\irrep{string}}{s=3/2}
\end{equation}
\end{guess}
Here $l_M$ is the eleven-dimensional Planck length, $V_{d+1}=R_s V_d$ the
volume of the M-theory torus $T^{d+1}$. The quantity $V_{d+1}/l_M^9=l_P^{d-8}$
is the U-duality invariant gravitational constant
in dimension $D=10-d$.
As an immediate check, the proposal has the appropriate scaling
dimension $d+1-9+3\times 2$ for an $R^4$ coupling in dimension
$D=10-d$.

\subsection{Strings, particles and membranes}

Before showing how this conjecture reproduces the tree-level and
one-loop terms, a few remarks are in order.
Firstly, our claim reduces to the Green-Gutperle conjecture \eqref{r4iib}
in the $d=1$ case of M-theory on $T^2$, or equivalently D=10 type IIB;
it also contains the $D=7,8$ extension of \cite{Kiritsis:1997em}
where the string multiplet transforms as a \irrep{(3,1)} and \irrep{5} of
$SL(3,\Zint)\times SL(2,\Zint)$ and $SL(5,\Zint)$ respectively, as
well as the $D=6$ proposal in \cite{Kiritsis:1997em}, although in
a refined way, since it is now a {\it constrained}
Eisenstein series that is involved. This is needed to obtain an
eigenmode of the Laplacian on the scalar manifold
$K\backslash \exc(\Real)$. Although such a requirement was strictly
proved in $D\ge 8$ \cite{Pioline:1998mn,Green:1998by}, it should
very plausibly hold in lower dimensions. Using the general formula
\eqref{eisgen}, we can compute the eigenvalue of the $\exc(\Zint)$
Eisenstein series in the string, particle and membrane
representations. These representations correspond to the leftmost,
rightmost and upmost nodes in the Dynkin diagram \eqref{dynkin} and
can be labelled by $SL(d+1)$ charges as follows: The
charges of the string multiplet are given in \eqref{tension}
while the particle and membrane multiplet have
charges $m_{[1]},m^{[2]},m^{[5]},m^{[1;7]} \ldots$
and $m,m^{[3]},m^{[1;5]} \ldots $ respectively and are listed
in Tables \ref{partm} and \ref{memm}.
Using the weights given in \cite{Obers:1998fb}  we can
compute the eigenvalues under the Laplacian:
\begin{subequations}
\label{psmeig}
\begin{eqnarray} \Delta_{\exc}
\eis{\exc(\Zint)}{string}{s}&=& \frac{s (4s-d^2+d-4)}{8-d}
\eis{\exc(\Zint)}{string}{s} \label{psmeig1} \\ \Delta_{\exc}
\eis{\exc(\Zint)}{particle}{s}&=& \frac{s (2(9-d)s
+d^2-17d+12)}{2(8-d)} \eis{\exc(\Zint)}{particle}{s}\\
\Delta_{\exc} \eis{\exc(\Zint)}{membrane}{s}&=& \frac{(d+1) s (
2s-3d + 4)}{2(8-d)} \eis{\exc(\Zint)}{membrane}{s}
\end{eqnarray}
\end{subequations}
(See Appendix \ref{slaped}, Eq. \eqref{laped} for the explicit
form of the Laplacian on the $K \backslash \exc (\Real)$ scalar
manifold). Substituting $s=3/2$ in \eqref{psmeig1} and noting that
the U-duality invariant factor $V_{d+1}/l_M^9=l_P^{d-8}$ is inert
under the Laplacian, we obtain
\begin{corol}
$R^4$ couplings in M-theory compactified on a torus $T^{d+1}$, $d\leq 7$ and $d\neq 2$, are
eigenmodes of the Laplacian on the symmetric space
$K\backslash\exc(\Real)$, with eigenvalue
\begin{equation}
\label{npeig}
\Delta_{\exc} f_{R^4} = \frac{3(d+1)(2-d)}{2(8-d)} f_{R^4}\ .
\end{equation}
\end{corol}
For $d=2$, this formula does not apply, due to the harmonic anomaly \eqref{harmanom}.  
Property \eqref{npeig} could in principle be proved from supersymmetry arguments
along the lines of \cite{Pioline:1998mn,Green:1998by}, and holds order
by order in the the weak coupling expansion. In particular, the tree
level contribution $V_d/(g_s^2 l_s^2) = e ^{\frac{12\phi}{d-8}}/l_P^{2-d}$,
albeit not U-duality invariant,
is an eigenmode of $\Delta_{\exc}$ with the same eigenvalue
as above, see Appendix \ref{slaped}, Eq. \eqref{lapedt}.

\begin{table}[ht]
\begin{center}
\begin{tabular}{|c|c||c|c|l|l|}
\hline
$D$&$d+1$ & U-duality group & irrep & $SL(d+1)$ content &
$SO(d,d)$ content \\ \hline
10& 1 &1& \irrep{1}  & \irrep{1}& \irrep{1}\\
9 & 2 &$SL(2,\Zint)$& \irrep{3}& \irrep{2}+ \irrep{1}& \irrep{2}+ \irrep{1}\\
8 & 3 &$SL(3,\Zint)\times SL(2,\Zint)$& \irrep{(3,2)}& \irrep{\bar{3}}+ \irrep{3}& \irrep{4}+ \irrep{2} \\
7 & 4 &$SL(5,\Zint)$& \irrep{10}& \irrep{\bar{4}}+ \irrep{6}& \irrep{6}+ \irrep{4}\\
6 & 5 &$SO(5,5,\Zint)$& \irrep{16}& \irrep{\bar{5}}+ \irrep{10}+ \irrep{1}& \irrep{8_V}+ \irrep{8_C}\\
5 & 6 &$E_{6(6)}(\Zint)$& \irrep{27}& \irrep{\bar{6}}+ \irrep{15}+
\irrep{6}& \irrep{10}+ \irrep{16}+ \irrep{1}\\
4 & 7 &$E_{7(7)}(\Zint)$& \irrep{56}& \irrep{\bar{7}}+ \irrep{21}+
\irrep{\bar{21}}+\irrep{7}& \irrep{12}+ \irrep{32}+ \irrep{12}\\
\hline
\end{tabular}
\end{center}
\caption{Particle multiplets of $\exc(\Zint)$\label{partm}}
\end{table}

\begin{table}[ht]
\begin{center}
\begin{tabular}{|c|c||c|c|l|l|}
\hline
$D$&$d+1$ & U-duality group & irrep & $SL(d+1)$ content &
$SO(d,d)$ content \\ \hline
10& 1 &1& \irrep{1}  & \irrep{1}& \irrep{1}\\
9 & 2 &$SL(2,\Zint)$& \irrep{1}& \irrep{1}& \irrep{1}\\
8 & 3 &$SL(3,\Zint)\times SL(2,\Zint)$& \irrep{(1,2)}& \irrep{1}+ \irrep{1}& \irrep{2} \\
7 & 4 &$SL(5,\Zint)$& \irrep{\bar 5}& \irrep{1}+ \irrep{4}& \irrep{4}+ \irrep{1}\\
6 & 5 &$SO(5,5,\Zint)$& \irrep{\bar{16}}& \irrep{1}+ \irrep{10}+ \irrep{5}& \irrep{8_C}+ \irrep{8_V}\\
5 & 6 &$E_{6(6)}(\Zint)$& \irrep{78}& \irrep{1}+ \irrep{20}+ \irrep{36}+\dots& \irrep{16}+ (\irrep{1}+\irrep{45})+\irrep{\bar{16}}\\
4 & 7 &$E_{7(7)}(\Zint)$& \irrep{912}& \irrep{1}+ \irrep{35}+\dots& \irrep{32}+ \dots\\
\hline
\end{tabular}
\end{center}
\caption{Membrane multiplets of $\exc(\Zint)$\label{memm}}
\end{table}

Secondly, we assumed according to conjecture 1 that the
Eisenstein series in the spinor of $SO(d,d,\Zint)$ reproduces
the one-loop threshold; for $d=1,2$, this is incorrect, since
we need also the conjugate spinor. However, the two contribute
to two different kinematic structures $(t_8 t_8 \pm
\epsilon_8\epsilon_8/4) R^4$, and \eqref{cr4} is only concerned
with the $+$ structure, while the $-$ is given at one-loop only
by the $SO(d,d,\Zint)$ Eisenstein series of order $s=1$ in the
conjugate spinor representation, and is U-duality invariant
by itself.

Thirdly, we could have considered the representation \eqref{c2} of
the one-loop threshold in terms of the Eisenstein series of order
$d/2-1$ in the {\it vector} of $SO(d,d,\Zint)$; the latter appears
as the leading term in the branching of the {\it particle}
multiplet of $\exc(\Zint)$ into representations of $SO(d,d,\Zint)$
(see Table \ref{partm}), so we would be led to the $\exc(\Zint)$
Eisenstein series of order $d/2-1$ in the particle representation.
Upon weak coupling expansion, this would start as a one-loop term
$\eis{SO(d,d,\Zint)}{\vect}{s=d/2-1}$ as in \eqref{c2}, but would
also include another perturbative term after Poisson resumming on
the vector charges, which would plausibly be the tree-level term
in \eqref{pertr4}. Similarly, we might have started from the
representation of the one-loop coupling in terms of the
$SO(d,d,\Zint)$ Eisenstein series of order 1 in the {\it
conjugate} spinor representation; the latter arises as the leading
term in the branching of the {\it membrane} multiplet of
$\exc(\Zint)$ into representations of $SO(d,d,\Zint)$ (see Table
\ref{memm}), so we would be led to the $\exc(\Zint)$ Eisenstein
series of order $1$ in the {\it membrane} representation, yielding
the correct one-loop term plus an extra (presumably tree-level)
perturbative contribution. Indeed, it is easy to check that the
Eisenstein series
\begin{subequations}
\begin{gather}
\eis{\exc(\Zint)}{string}{s=3/2}\ ,\quad
\eis{\exc(\Zint)}{particle}{s=d/2-1}\ ,\quad
\eis{\exc(\Zint)}{membrane}{s=1}\ ,\quad \\
\eis{\exc(\Zint)}{string}{(d+1)(d-2)/4}\ ,\quad
\eis{\exc(\Zint)}{particle}{s=3(d+1)/(9-d)}\ ,\quad
\eis{\exc(\Zint)}{membrane}{s=3(d-2)/2}\ ,\quad
\end{gather}
\end{subequations}
are all degenerate with $f_{R^4}$ under the Laplacian. It is thus
quite tempting to conjecture
\begin{guess}
The Eisenstein series of $\exc(\Zint)$, $d>2$, in the string multiplet
representation at the particular order $s=3/2$ is equal to
the one in the particle multiplet of order $s=d/2-1$, and to
the one in the membrane multiplet of order $s=1$,
up to numerical coefficients and powers of Newton's constant:
\begin{equation}
\label{ccc}
\frac{V_{d+1}}{l_M^9}\eis{\exc(\Zint)}{\irrep{string}}{s=3/2}
=\frac{\Gamma(d/2-1)}{\pi ^{d/2-2}}\eis{\exc(\Zint)}{particle}{s=d/2-1}=
\frac{V_{d+1}}{l_M^9}\eis{\exc(\Zint)}{\irrep{membrane}}{s=1}
\end{equation}
\end{guess}
Again, it is easy to check that the scaling dimensions match. Note
that the restriction $d>2$ applies because we are making use of
\eqref{c1}. For $d=3,4$, this conjecture nicely checks with
\eqref{slinv},\eqref{c0},\eqref{c1}:
\begin{subequations}
\begin{gather}
\eis{SL(5,\Zint)}{\irrep{5}}{s=3/2}=
\pi \eis{SL(5,\Zint)}{\irrep{10}}{s=1/2}=
\eis{SL(5,\Zint)}{\irrep{\bar 5}}{s=1} \\
\eis{SO(5,5,\Zint)}{\irrep{10}}{s=3/2}=
\eis{SO(5,5,\Zint)}{\irrep{16}}{s=1}=
\eis{SO(5,5,\Zint)}{\irrep{\bar {16}}}{s=1}
\end{gather}
\end{subequations}
up to factors of Newton's constant, whereas $d>4$ gives new
identities. The automorphic forms in \eqref{ccc}
should give three different representations of the
same $R^4$ threshold in M-theory on $T^{d+1}$.

\subsection{Weak coupling expansion and instanton effects}
Now, in order to justify the claim \eqref{cr4}, we need to
show that it reproduces the perturbative contributions in
\eqref{pertr4} in a weak coupling expansion. This is achieved
as usual by a sequence of Poisson resummations on the integral
representation
\begin{multline}
\label{sonn} f_{R^4}= \frac{V_d}{g_s^2} \frac{\pi ^s}{\Gamma(s)}
\int\int  \frac{dt d\theta }{t^{1+s}} \hat{\sum} \\ \exp
\left\{-\frac{\pi} {t} \left[ m^2 + \frac{1}{g_s^2} \left[ (\tilde
m^i)^2 + V_d^2 (m_i)^2 \right]+ \frac{V_d^2}{g_s^4} \bar
m^2\right] +2\pi i \theta \left( m\bar m + m^i m_i \right)
\right\}
\end{multline}
where the integral runs from 0 to $+\infty$ for $t$ and $0$ to $1$
for the Lagrange multiplier $\theta$; the sum is on unrestricted
integers, not vanishing all at the same time, and for definiteness
we restricted to the $d=4$ case with vanishing RR fields, and
defined $m_i=\epsilon_{ijkl}m^{sjkl}/3!$ and $\bar
m=\epsilon_{ijkl}m^{ijkl}/4!$. The leading contribution as $g_s\to
0$ arises from the term $m_i=m^i=\bar m=0$ with $m\neq 0$, and
reproduces the tree-level term in \eqref{pertr4}. After
subtracting this term, the sum over $m$ is now unrestricted, and
we can Poisson resum on $m$ using the formula \eqref{pois}:
\begin{multline}
f_{R^4}=
2\zeta(2s)\frac{V_d}{g_s^2}
+\frac{V_d}{g_s^2}
\frac{\pi ^s}{\Gamma(s)}
\int\int  \frac{dt d\theta }{t^{1+s-\frac{1}{2}}}
\hat{\sum} \\
\exp \left\{ -\pi t (m+\theta \bar m)^2 -\frac{\pi}{t} \left[
\frac{1}{g_s^2} \left[ (\tilde m^i)^2 + V_d^2 (m_i)^2
\right]+ \frac{V_d^2}{g_s^4} \bar m^2\right]
+2\pi i \theta m^i m_i \right\}
\end{multline}
where we should substitute $s=3/2$.
This now contains several contributions when $\bar m=0$
(and therefore $m_i,m^i$ not simultaneously zero): for $m=0$,
we precisely recover the Eisenstein series of order $s-1/2=1$ in the
spinor representation, whereas $m\neq 0$ contains non-perturbative
$e^{-1/g_s}$ effects:
\begin{multline}
f_{R^4}=
2\zeta(2s)\frac{V_d}{g_s^2} +
\left(\frac{V_d}{g_s^2}\right)^{\frac{3}{2}-s}
\eis{SO(d,d,\Zint)}{\spi}{s-1/2}
+\left(\frac{V_d}{g_s^2}\right)^\frac{3-2s}{4}
\frac{2\pi ^s}{\Gamma(s)}
\hat{\sum_m} \hat{\sum_{m^i,m_i}} \delta(m_i m^i) \\
\left[ \frac{m^2 V_d}{(\tilde m^i)^2 + V_d^2 (m_i)^2} \right]
^\frac{2s-1}{4} K_{s-\frac{1}{2}}\left( -\frac{2\pi|m|}{g_s}
\sqrt{ (\tilde m^i)^2 + V_d^2 (m_i)^2 } \right)
+\dots
\end{multline}
Using the saddle point approximation \eqref{asbessel} of the
Bessel function at $s=3/2$, we see that these
non-perturbative terms can be interpreted
as superposition of Euclidean D0 and D2-branes wrapped on
a one-cycle $m^i$ or a three-cycle $\epsilon^{ijkl} m_l$ of $T^4$,
preserving half of the supersymmetries $(m^im_i=0)$
\cite{Pioline:1997pu}.
In addition to these terms,
we have further contributions arising from $\bar m\neq 0$,
\begin{multline}
\left(\frac{V_d}{g_s^2}\right)^\frac{3-2s}{4}
\frac{2\pi ^s}{\Gamma(s)} \int_0^1 d\theta
\hat{\sum_m} \hat{\sum_{m^i,m_i}}
\left[ \frac{m^2 g_s^2 V_d}{V_d^2 \bar m^2+ g_s^2 (\tilde m^i)^2 +
g_s^2 V_d^2 (m_i)^2} \right]^\frac{2s-1}{4}\\
K_{s-\frac{1}{2}}\left( -\frac{2\pi|m+\theta \bar m|}{g_s^2}
\sqrt{ V_d^2 \bar m^2 + g_s^2 (\tilde m^i)^2 + g_s^2 V_d^2 (m_i)^2 } \right)
e^{2\pi i \theta m^i m_i }
\end{multline}
which behave superficially as $e^{-1/g_s^2}$. Such non-perturbative
effects are certainly unexpected in toroidal compactifications to
$D>4$, since there are no half-BPS instanton configurations
with this action (the NS5-brane does have a tension scaling
as $1/g_s^2$, but it can only give rise to Euclidean configurations
with finite actions when $D\leq 4$). Unfortunately, the infinite
sum is not uniformly convergent ( $|m+\theta n|$ can vanish at any
rational value of $\theta$), so we cannot be positive about the
existence of such effects at that stage\footnote{One may
carry out the Gaussian $\theta$ integration by summing
over $m$ modulo $\bar m$ only and then compute the sum over $m$,
but this only takes us back to \eqref{sonn}.}. The matching
of the tree-level and one-loop contributions together with
the consistent interpretation of the D-brane contribution is
however a strong support to our conjecture.

\section[Higher genus integrals and higher derivative couplings]
{Higher genus integrals and higher derivative couplings
\label{highg}}
\subsection{Genus $g$ modular integral}
Having discussed the modular integrals arising in one-loop
amplitudes, one may ask if our methods carry over to higher-loop
amplitudes, which are notoriously difficult to evaluate. We shall
not attempt to make any full-fledged higher-genus amplitude
computation, but we will consider the higher-genus analogue of
\eqref{oneloop}, namely the integral of a lattice partition
function on the $3g-3$-dimensional moduli space of genus $g$
curves
\begin{subequations}
\label{gloop}\begin{gather} I^g_d=\int_{{\cal{M}}_g}  d\mu~
Z_{d,d}^g \left(g_{ij},B_{ij};\tau\right) \\ Z_{d,d}^g= V_d^{g}
\sum_{m^i_A,n^{iA}\in \Zint}\exp\left[ -\pi
(g_{ij}+B_{ij})(m^{i}_A+\tau_{AB}n^{iB}) \tau_2^{AC}
(m^j_C+\bar\tau_{CD} n^{jD})\right]
\end{gather}
\end{subequations}
Here, the integers $m^i_A, n^{iA}$ denote the winding numbers
along the cycles $\gamma_A$ and $\gamma ^A$ of a symplectic basis
of the homology lattice of the genus $g$ curve, and the period
matrix $\tau_{AB}$, of positive definite imaginary part, describes
the complex structure on the curve.  $(m^i_A, n^{iA})$ transforms
as a symplectic vector under $Sp(g,\Zint)$ which now plays the
role of the modular group. $\mu$ is the modular invariant
Weil-Peterson measure on the moduli space ${\cal{M}}_g$ of genus
$g$ curves (see for instance \cite{D'Hoker:1988ta} for a review).
Except for $g=1,2$, $\tau_{AB}$ is a redundant parametrization of
the Teichm{\"u}ller space of dimension $(3g-3)$, constrained by
Schottky relations. Nonetheless, for our computation it will be
convenient to consider it as a set of independent parameters
living in the symmetric space $U(g)\backslash Sp(g,\Real)$, with
partial Iwasawa decomposition
\begin{equation}
\label{mosp} M(\vect)  = \begin{pmatrix}  \mathbb{I}_g &  \\
\tau_1 & \mathbb{I}_g
\end{pmatrix} \cdot
\begin{pmatrix} \tau_2^{-1} & \\ & \tau_{2} \end{pmatrix} \cdot
\begin{pmatrix} \mathbb{I}_g & \tau_{1} \\ & \mathbb{I}_g \end{pmatrix}
\end{equation}
Note that the boost parameter $\tau_1$ is now symmetric, as
imposed by the symplectic condition. From this it is
straightforward (see Appendix \ref{lap} for the derivation)  to
determine an $Sp(g,\Real)$ invariant second order differential
operator, namely the Laplacian on this manifold:\footnote{Again,
the derivatives w.r.t. to the symmetric matrices $\tau_1$
and $\tau_2$ are computed in terms of the diagonally rescaled
matrices $(1-\delta_{AB}/2)\tau_{1,2;AB}$.}
\begin{equation}
\label{splap}
\Delta_{Sp(g)}=
\frac{1}{4}\tau_{2AC} \tau_{2BD}
\left( \frac{\partial}{\partial \tau_{1AB}}
\frac{\partial}{\partial \tau_{1CD}}
+\frac{\partial}{\partial \tau_{2AB}}
\frac{\partial}{\partial \tau_{2CD}} \right)
\end{equation}
which reduces to twice the $SL(2,\Real)$ Laplacian \eqref{lsl2} for
$g=1$. An explicit computation along the same lines as before
shows that the genus $g$ lattice sum continues to obey a partial
differential equation
\begin{equation}
\left[ \Delta_{SO(d,d)} - \Delta_{Sp(g)} +\frac{dg(d-g-1)}{4}
\right] Z_{d,d}^{g} =0
\end{equation}
The non-trivial step is now to integrate by parts the
$\Delta_{Sp(g)}$ term. As we already emphasized, except in the
$g=1,2$ case, the integration measure is {\it not} the
$Sp(g,\Real)$-invariant measure on $\tau$-space, but its
restriction to the solution of Schottky constraints. Nevertheless,
we assume that the expression of $\Delta_{Sp(g)}$ in terms of the
independent coordinates still yields the appropriate Laplacian,
and we can therefore integrate it by parts. Under this plausible
assumption, we obtain
\begin{equation}
\label{eighg}
\Delta_{SO(d,d)} I_d^g = \frac{dg(g+1-d)}{4} I_d^g
\end{equation}
Quite amazingly, comparison with \eqref{soeigval} shows that this
eigenvalue agrees with the order $s=g$ Eisenstein series in the
spinor and conjugate spinor representation. We are therefore led
to the

\begin{guess} The integral \eqref{gloop} of the $(d,d)$ lattice
partition function on the fundamental domain of the moduli space of
genus $g$ Riemann surfaces is given, up to an overall
factor, by the $SO(d,d,\Zint)$
Eisenstein series of order $g$ in the spinor representation:
\begin{equation}
\label{c3}
I_d^g \propto \eis{SO(d,d,\Zint)}{\spi}{s=g} + \eis{SO(d,d,\Zint)}{\spb}{s=g}
\end{equation}
\end{guess}
Note that the superposition of the two spinor representations is
required by the $O(d,d,\Zint)$ invariance of the integrand.
Normalizing \eqref{c3} would require a knowledge of the
Weil-Peterson volume of the moduli space of genus $g$ curves.

\subsection{$N=4$ topological string and higher derivative terms}

The conjecture \eqref{c3} is less substantiated than the 1-loop
conjecture \eqref{c1}, since
we do not have a second differential operator at our disposal, nor can we
control the large volume limit of the lefthand side of \eqref{c3}.
It is however
strongly reminiscent of the genus $g$ partition function of the
$N=4$ topological string \cite{Berkovits:1995vy} on $T^2$,
which was shown to be exactly given by
the Eisenstein series of order $s=g$ in
the spinor representation $\eis{SL(2,\Zint)}{2}{s=g}(T)$~\cite{Ooguri:1995cp}.
The precise result
\begin{equation}
\label{ova}
F^g(u_L,u_R) \propto \hat{\sum_{(m,n)}}
|n+mT|^{2g-4}\left(\frac{u_L^+u_R^+}{n+mT}
+\frac{u_L^-u_R^-}{n+m\bar T} \right)^{4g-4}
\end{equation}
involves a set of harmonic variables $u$, with charge 1/2 under
the R-symmetry $SO(2)$. This result was obtained from a set of
first-order differential equations, which, loosely speaking, are
nothing but the holomorphic half of our second-order differential
equation \eqref{eighg}. It was subsequently used to derive a set
of higher derivative topological couplings $R^4 H^{4g-4}$ in type
IIB string compactified over $T^2$ \cite{Berkovits:1998ex}. Our
conjecture \eqref{c3} suggests a natural generalization of these
results to lower dimensions, which we shall now present.

The topological amplitude \eqref{ova} can be identified with
higher derivative couplings $R^4 H^{4g-4}$ in type IIB string theory
on $T^2$ in the following way. The field-strength of the Ramond
two-forms  ${\cal{B}}_{\mu\nu},
{\cal{D}}_{\mu\nu 12}$ transform as a doublet $H_{RR}^i$ of $SL(2,\Real)_T$.
Using the $SO(2)\backslash SL(2,\Real)$ two-bein $e_i^{\pm\pm}$, these two
three-forms
can be converted into an $SO(2)$ doublet $H_{RR}^{\pm\pm}=
H_{RR}^i e_i ^{\pm\pm}$, and
further contracted with the harmonic variables into an $SO(2)$
invariant $\hat H_{RR}=u_L^+ u_R^+ H_{RR}^{--} + u_L^- u_R^- H_{RR}^{++}$.
Integrating \eqref{ova} against $R^4 \hat H^{4g-4}$ in harmonic
superspace\footnote{The precise contraction of the Lorentz indices
is also obtained by dressing $\hat H_{RR}$ with Grassmann parameters,
and generalizes the usual $t_8t_8+\epsilon_8\epsilon_8/4$
combination \cite{Berkovits:1998ex}.} yields the physical coupling
\begin{equation}
\int d^8 x \sqrt{-\gamma} \sum_{p=2-2g}^{2g-2} (-)^p
R^4 (H^{++}_{RR})^{2g-2+p} (H_{RR}^{--})^{2g-2-p}
\hat{\sum_{m,n}} \frac{{T_2}^g}
{(m+nT)^{g+p} (m+n\bar T)^{g-p}}
\end{equation}
Using the identity $(m+nT)H_{RR}^{--}-(m+n\bar T)H_{RR}^{++}=
m H_{RR}^2 -n H_{RR}^1$,
we can rewrite the above result in the more suggestive way
\begin{equation}
\label{obv} \int d^8 x \sqrt{-\gamma} \ \hat{\sum_{m,n}} \
\frac{R^4 \ (m_i H_{RR}^i)^{4g-4}} {\left(m_i
{M}^{ij}(\irrep{\spb}) m_j\right)^{3g-2}}
\end{equation}
where $M(\irrep{\spb})$ is the mass matrix in the conjugate spinor
representation $\spb$ of $SO(2,2,\Zint)$. Indeed, $H^i$ transforms
as a conjugate spinor under the T-duality group, while
$m_i=(m,n)$ transforms in the dual way. More generally,
in type IIB on $T^d$ the 2-form and 1-form potentials in the RR sector
transform in the conjugate spinor and spinor representation of $SO(d,d)$
respectively, while in type IIA these two representations are interchanged.

Using the representation \eqref{obv}, the generalization of the
$g$-loop $R^4 H^{4g-4}$ coupling to lower dimensions is then
obvious: in type IIA variables,
\begin{guess}
The $R^4 H^{4g-4}$ couplings between 4 gravitons and $4g-4$
Ramond three-form  field-strengths in type IIA compactified on $T^d$, $d\leq 4$
are given at genus $g$ by the $SO(d,d,\Zint)$
constrained Eisenstein series in the spinor representation
with insertions of $4g-4$ charges:
\begin{equation}
\label{rhp} I=\int d^{10-d} x \sqrt{-\gamma}\ \hat{\sum_{m}}
\delta(m\wedge m)\ e^{6(g-1)\phi} \ \frac{R^4 \ (m\cdot
H_{RR})^{4g-4}} {\left(m\cdot {M}(\irrep{\spi}) \cdot
m\right)^{3g-2}}
\end{equation}
\end{guess}
where $\phi$ is the T-duality invariant dilaton, related to the
ten-dimensional coupling as $e^{-2\phi}=V_d/g_s^2 l_s^d$, and we
work in units of $l_s$. The restriction $d\leq 4$ is due to the
fact that for $D = 5$ three-form field-strengths are Poincar{\'e}
dual to two-form field-strengths, while for $D=4$ they become part
of the scalar manifold after dualization. A similar conjecture
also holds for the coupling computed by the topological B-model
\cite{Berkovits:1995vy},
\begin{guess}
The $R^4 F^{4g-4}$ couplings between 4 gravitons and $4g-4$ Ramond
two-form field-strengths
in type IIA compactified on $T^d$, $d\leq 6$  are given at genus $g$
by the $SO(d,d,\Zint)$ constrained Eisenstein series in the
conjugate spinor representation with insertions of $4g-4$ charges:
\begin{equation}
\label{rfp}
I=\int d^{10-d} x \sqrt{-\gamma} \
\hat{\sum_{m}} \delta(m\wedge m) \
e^{6(g-1)\phi} \frac{R^4 \ (m\cdot F_{RR})^{4g-4}}
{\left(m\cdot {M}(\irrep{\spb}) \cdot m\right)^{3g-2}}
\end{equation}
\end{guess}
Here, the restriction $d\leq 6$ is due to the fact that for $D =
3$ two-forms field-strengths become part of the scalar manifold
after Poincar{\'e} dualization. The relation between these two
conjectures and the genus $g$ integral \eqref{c3} is similar to
the case of $(t_8t_8\pm\epsilon_8\epsilon_8/4)R^4$ couplings in
dimensions 8 or higher: the insertions of the vertex operators of
the four gravitons and the $4g-4$ two-forms $F_{RR}$ or
three-forms $H_{RR}$ saturate the fermionic zero-modes and select
one out of the two spinor contributions in the modular integral
\eqref{c3}. The end results \eqref{rhp} and \eqref{rfp} involve
covariant modular functions instead of invariant ones, but behave
as Eisenstein series of order $3g-2-(4g-4)/2=g$ for most purposes.
They generalize the $SL(2,\Zint)$ modular functions
$f^{p,q}={\hat{\sum}}
\tau_2^{(p+q)/2}/[(m+n\tau)^p(m+n\bar\tau)^q]$ invariant up to a
phase, that were also used in the context of non-perturbative type
IIB string in \cite{Kehagias:1997cq,Green:1998me}.

\subsection{Non-perturbative $R^4 H^{4g-4}$ couplings}
Having put the $g$-loop amplitude in a manifestly T-duality
invariant form \eqref{rhp}, it is now straightforward to propose a
non-perturbative completion, invariant under the full U-duality
group. For that purpose, we note that the set of three-form
field-strengths in M-theory compactified on $T^{d+1}$ fall into a
representation of {\exc} dual to the string multiplet which
already appeared in Section \ref{npert} (this is strictly speaking
only correct for $D \geq 5$ as explained below \eqref{rhp}). The
string multiplet decomposes under $SO(d,d,\Zint)$ into a singlet
(the Neveu-Schwarz $H_{NS}$) a spinor (the Ramond  three-forms
obtained by reducing the M-theory four-form field-strength), as
well as further terms for $d\ge 4$. It is therefore tempting to
propose
\begin{guess}
The $R^4 H^{4g-4}$ couplings between 4 gravitons and $4g-4$
three-form field-strengths in M-theory compactified on $T^{d+1}$,
$d\leq 4$ are exactly given, up to a power of Newton's constant,
by the $\exc(\Zint)$ constrained Eisenstein series in the string
representation with insertions of $4g-4$ charges:
\begin{equation}
\label{rh}
I=\frac{V_{d+1}}{l_M^9} \int d^{10-d} x \sqrt{-\gamma}\
\hat{\sum_{m}} \delta(m\wedge m)\
\frac{R^4\ (m\cdot H)^{4g-4}}
{\left(m\cdot {M}(\irrep{string}) \cdot m\right)^{3g-\frac{3}{2}}}
\end{equation}
\end{guess}
As an immediate check, we note that this proposal has the appropriate
scaling dimension. The leading contribution arises by restricting
the summation to $m^s\neq 0$ only, where $m^s$ is the top charge
in the string multiplet $m$, contracted with the top three-form
$H^{NS}$:
\begin{equation}
I=\frac{V_{d}}{g_s^2 l_s^8} \int d^{10-d} x \sqrt{-\gamma}\
2\zeta\left(2g+1\right)
\ R^4\ H_{NS}^{4g-4} + \dots
\end{equation}
corresponding to a tree-level interaction involving the
Neveu-Schwarz three-form only. The next-to-leading contribution is
obtained by Poisson resummation on the integer $m^s$, and setting
the dual integer to zero, as in our analysis of $R^4$ couplings.
This has the effect of setting $m\cdot H=m_{RR}\cdot H_{RR}$ (for
vanishing value of the Ramond scalars) and shifting the order
$3g-3/2\to 3g-2$. We thus reproduce the $g$-loop result
\eqref{rhp}. The analysis of non-perturbative effects is as in the
$R^4$ case, and shows order $e^{-1/g_s}$ D-brane effects as well
as, for $d\ge 4$, contributions superficially of order $e^{-1/g_s^2}$ . More
explicitly, in the simplest example of ten-dimensional type IIB
theory, we obtain, in units of the 10D Planck length,
\begin{multline}
\hat{\sum_{m,n}}
\left[\frac{\tau_2}{|m+n\tau|^2}\right]^{3g-\frac{3}{2}}
\ R^4(m H_{NS}-n H_{RR})^{4g-4}\
=2\zeta\left(2g+1\right)
R^4H_{NS}^{4g-4} \\
+ 2\sqrt{\pi} \tau_2^{\frac{5}{2}-3g}
\frac{\Gamma(3g-2)}{\Gamma(3g-\frac{3}{2})} \zeta(6g-4)
\ R^4(H_{RR}-\tau_1 H_{NS})^{4g-4}
+ O(e^{-1/g_s})
\end{multline}

Turning finally to the case of non-perturbative $R^4 F^{4g-4}$
couplings, we note that the two-form field-strengths of
M-theory compactified on $T^{d+1}$ transform as the dual
of the particle multiplet. The particle multiplet is the representation
associated to the rightmost node in the Dynkin diagram \eqref{dynkin} (see
Ref. \cite{Obers:1998fb} for further details).
It is thus quite natural to propose
a non-perturbative completion as
\begin{equation}
\label{rf}
I=\int d^{10-d} x \sqrt{-\gamma}\
\hat{\sum_{m}} \delta(m\wedge m)\
\frac{R^4\ (m\cdot F)^{4g-4}}
{\left(m\cdot {M}(\irrep{particle}) \cdot m\right)^{4g-5+\frac{d}{2}} }
\end{equation}
where the power $4g-5+d/2$ has been set by dimensional analysis.
The particle multiplet decomposes as a vector and conjugate spinor
of $SO(d,d,\Zint)$ in that order, so that this proposal implies a
one-loop term given by the $SO(d,d,\Zint)$ Eisenstein series of
order $2g-3+d/2$ in the vector representation, plus a higher
perturbative term which should reproduce the genus $g$ term
\eqref{rfp}. Due to the presence of constraints, we are
unfortunately not able to prove this statement at present. For
$g=1$, this conjecture is implied by the alternative form of the
$R^4$ threshold in \eqref{ccc}. Note that this proposal may in
principle lift the difficulty raised by Berkovits and Vafa, who
noted that in 8 dimensions the non-perturbative generalization of
the genus $g$ $R^4 F^{4g-4}$ terms should include a mixing between
the $U(1)\backslash SL(2,\Real)$ and $SO(3)\backslash SL(3,\Real)$
moduli \cite{Berkovits:1998ex}. Here the mixing is built-in since
the particle multiplet transforms in the $\irrep{(3,2)}$ of
$SL(3,\Zint)\times SL(2,\Zint)$. Let us finally note that our
techniques could also be used to generalize the conjectures about
$\nabla^{2k} R^4$ and $R^{3m+1}$ terms
\cite{Russo:1997mk,Kehagias:1997jg}, but the status of these is
less clear.

\section{Conclusions}
Duality provides strong constraints on the non-perturbative
extension of string theory. It is especially powerful in
vacua with many supersymmetries, where physical amplitudes
and low energy couplings have to be invariant under the
symmetry group. For a restricted class of BPS saturated
couplings, the supersymmetry constraints close into
a set of partial differential equations, which together
with perturbative boundary conditions allows to determine
the result exactly. Such techniques have enabled us
to obtain convenient representations of one-loop thresholds
manifestly invariant under T-duality, to compute higher-genus
amplitudes not tractable otherwise, and to propose
an exact non-perturbative completion of $R^4 H^{4g-4}$ couplings
in toroidal compactifications of M-theory. Upon expansion
in weak coupling, these results reveal a tree-level
and $g$-loop contribution, non-perturbative order $e^{-1/g_s}$ effects
that can be attributed to Euclidean D-branes wrapped on various
cycles of the internal torus, as well as further ill-understood
non-perturbative effects superficially of order $e^{-1/g_s^2}$, appearing
in dimension $D=6$ and lower.
It would be very interesting to ascertain the behaviour of
these effects, and eventually give an instantonic interpretation
for them.  In $D=4$ we expect such $e^{-1/g_s^2}$ effects from
the Euclidean NS5-brane wrapped on $T^6$ which should be extracted from
our conjecture \eqref{cr4}. Finally, the generalization to $D\leq 2$
should involve Eisenstein-like series for affine Lie algebras
and even hyperbolic Kac-Moody algebras.

We have focused in this work on half-BPS saturated couplings in
maximally supersymmetric theories. It would be interesting to
extend our techniques to (i) couplings preserving a lesser amount
of supersymmetry, and (ii) half BPS states in theories with less
supersymmetry. Given that the quadratic half-BPS constraint imposes
second order differential equations and that the quarter-BPS condition
is cubic in the charges,
one may envisage that quarter-BPS saturated couplings
should be eigenmodes of a cubic Casimir operator, and expressable as
generalized Eisenstein series.
As for the second issue, one has to face
  situations
where the gauge symmetry can be enhanced at a particular point
in the moduli space, a case where Eisenstein series seem to be of little
relevance. The differential equations \eqref{sonkeq} and
the generalized prepotentials of \cite{Kiritsis:1997hf} should prove useful
for constructing automorphic forms with the required singularity structure,
generalizing \cite{Mayr:1995rx,Harvey:1996fq,Borcherds:1996,Berglund:1997eb}.
Particularly interesting cases include the toroidal
compactifications of the heterotic string, where five-brane
instantons are little understood; type IIB compactified
on $K_3$, where the moduli space unifies the dilaton with
the other scalars in a simple form $[SO(5)\times SO(21)]\backslash
SO(5,21)$ and where tensionless strings appear at singularities
of $K_3$; the FHSV model \cite{Ferrara:1995yx}, where the duality group is
broken to a subgroup of $SO(2,10,\Zint)$ by the freely acting orbifold
construction.

On a more mathematical level, our results provide a wealth of
explicit examples of modular functions on symmetric spaces of
non-compact type $K\backslash G$, with $G$ a real simply laced Lie
group in the normal real form, that generalize the Eisenstein
series on the fundamental domain of the upper half-plane. These
functions can be associated to any fundamental representation of
$G$, and are eigenmodes of the Laplacian with an easily computable
eigenvalue. From analyzing their asymptotics and their behaviour
under the Laplace operator as well as some other differential
operator, we have been able to obtain a number of relations
between Eisenstein series in various representations, although we
had to content ourselves with conjectures rather than proofs in
several cases. This has shown that Eisenstein series may become
equal for certain values of the order $s$, the most useful example
being the equality of the vector, spinor and conjugate spinor
Eisenstein series of $SO(d,d,\Zint)$ at $s=d/2-1$, $s=1$ and $s=1$
respectively. On the other hand, two Eisenstein series with the
same eigenvalue under the Laplacian may still be separated by an
extra differential operator, like $\square_d$ in the $SO(d,d)$
case. We have not addressed the question of the analyticity of
Eisenstein series with respect to the order $s$: this would
require an asymptotic expansion analogous to \eqref{eisfour} or
\eqref{expsl} with a uniformly suppressed general term.
Unfortunately, it seems that the presence of constraints tends to
give rise to ill-behaved expansions such as \eqref{largevolV}.
This problem is the mathematical counterpart of the physical one
raised above, namely understanding the instanton effects that are
superficially of order $e ^{-1/g_s^2}$. It would be interesting to
understand more precisely what Eisenstein series are needed to
generate the spectrum of the Laplace operator for any eigenvalue
(note in that respect that the order $s$ is no longer a good
parametrization, since the relation between the eigenvalue and $s$
depends on the representation). From a mathematical point of view
however, Eisenstein series are the least interesting part of the
spectrum on such manifolds, which should also include a discrete
family of cusp forms. Perhaps string theory will provide an
explicit example of these elusive objects.


\vskip 5mm

\noindent {\it Acknowledgements:}~We are grateful to N.
Berkovits, J. Bernstein, R. Borcherds, G. W. Moore, M.
Petropoulos, E.Verlinde, J-B. Zuber and G. Zwart
for useful discussions or correspondence, and especially to E.
Kiritsis for participation at an early stage of this project. B.P.
thanks Nordita and both of the authors the CERN Theory Division
for their kind hospitality and support during the completion of
this work.

\vskip 5mm

\noindent {\it Note added (Jan. 2010):}~We have corrected misprints in Eqs. 2.9, 2.11, 2.19, 4.3, and 
an error in the derivation in Sec. C.1 and C.3  of the large volume expansion of the Eisenstein 
series in the vector representation of $SO(d,d)$, and in the spinor representations of $SO(4,4)$.
This mistake does not affect the result for $s=d/2-1$, which is the case relevant for threshold
integrals. We also removed an erroneous conjecture below Eq. 2.12, and added a comment 
(footnote 4) about the need to regulate the integral (3.16).

\vskip 1cm
\centerline{\bf \large Appendices}

\appendix

\section{$Gl(d)$, $SL(d)$, $SO(d,d)$ and $Sp(g)$ Laplacians \label{lap}}

In this appendix we give some details of the derivation of the
Laplacians \eqref{gllap}, \eqref{sllap}, \eqref{solap} and \eqref{splap}
on the
scalar manifolds for the four cases of $Gl(d)$, $SL(d)$, $SO(d,d)$
and $Sp(g)$ symmetry, as well as some useful alternative forms.
The Laplacians are  computed from the general expression
\begin{equation}
\label{metric}
\Delta = \frac{1}{\sqrt{\gamma}}\partial_\mu \sqrt{\gamma} \gamma
^{\mu\nu} \pa_\nu \ ,\qquad
ds^2=\gamma_{\mu\nu}dx^\mu dx^{\nu}=-\frac{1}{2}
\Tr\left(dM dM^{-1}\right)
\end{equation}
where $\gamma$ is the bi-invariant metric on the symmetric space
$K\backslash G$, parametrized by the symmetric matrix $M$.

\subsection{Laplacian on the $SO(d)\backslash Gl(d,\Real)$ and
$SO(d)\backslash SL(d,\Real)$ symmetric spaces} For the
$SO(d)\backslash Gl(d)$ case, we can choose $M=g$ a symmetric
positive definite matrix, and the metric $ds^2$ and volume element
take the form
\begin{equation}
ds^2 = g^{ik} g^{jl} dg_{ij} dg_{kl}\ ,\qquad
\label{gldet}
\det(ds^2)= 2^{\frac{d(d-1)}{2}} (\det g)^{-(d+1)}
\end{equation}
Its inverse is easily computed by ordering the indices,
\begin{equation}
ds_{\rm inv}^2 = \sum_{i,j} g_{ij}g_{ij} dg^{ii} dg^{jj} +
\frac{1}{2}\sum_{i<j;k<l} \left( g_{ik} g_{jl} + g_{il} g_{jk}
\right)
 dg^{ij} dg^{kl}
+  2 \sum_{i;k<l} g_{ik} g_{il} dg^{ii} dg^{kl}
\end{equation}
and using the relation
\begin{equation}
\frac{\partial  \det g}{\partial g_{ij}} = (2-\delta_{ij})
g^{ij} \det g\ .
\end{equation}
We find, after some algebra,
\begin{equation}
\Delta_{Gl(d)}=
\sum_{i\leq j;k\leq l} \frac{\partial}{\partial g_{ij}}
g_{ik} g_{jl}
\frac{\partial}{\partial g_{kl}}
- \frac{d+1}{2} \sum_{i\leq j} g_{ij}
\frac{\partial}{\partial g_{ij}}
\end{equation}
which can also be put in the form
\begin{equation}
\label{gllap2}
\Delta_{Gl(d) }=
\sum_{i\leq j;k\leq l}
g_{ik} g_{jl}
\frac{\partial}{\partial g_{ij}}
\frac{\partial}{\partial g_{kl}}
+ \frac{d+1}{2} \sum_{i\leq j} g_{ij}
\frac{\partial}{\partial g_{ij}}
\end{equation}
In order to avoid the cumbersome sums over ordered indices, it
is convenient to introduce the diagonally rescaled metric
\begin{equation}
\label{gtil}
\tilde{g}_{ij} = (1-\delta_{ij}/2 ) g_{ij}   \ .
\end{equation} This then satisfies
the properties
\begin{equation}
\frac{\pa g_{ij}}{\pa \tilde{g}_{kl} } =
\delta_i^k \delta_j^l  + \delta_i^l \delta_j^k \sp
\frac{\pa \det g}{\pa \tilde{g}_{ij} } = 2 g^{ij} \det g
\end{equation}
which allows to write the above Laplacian in the covariant form
\begin{equation}
\label{gllapa}
\Delta_{Gl(d)} =\frac{1}{4} g_{ik} g_{jl}
 \frac{\pa}{ \pa \tilde g_{ij}}
 \frac{\pa}{  \pa \tilde g_{kl} }
 + \frac{d+1}{4}  g_{ij} \frac{\pa}{ \pa \tilde g_{ij} }
\end{equation}
where now repeated indices are summed over without further restrictions.
This is the form given in \eqref{gllap}, where we omitted
the tilde on the redefined metric as done throughout the text of
the paper for simplicity of notation.

To compute the Laplacian on the $SO(d)\backslash SL(d)$ symmetric
space from this, we decompose the element $g$ of $Gl(d)$ as $g= t
\tilde g$, with $\det \tilde g=1$. The metric then takes the form
\begin{equation}
ds_{Gl}^2 = ds_{Sl}^2 + d \left(\frac{dt}{t}\right)^2 \ \ ,\qquad
\sum_{i\leq j} g_{ij} \frac{\partial}{\partial g_{ij}} =
t\partial_t
\end{equation}
so that the Laplacian reads
\begin{equation}
\label{glsl} \Delta_{Gl(d)} = \Delta_{SL(d)}+ \frac{1}{d}
t\partial_t t\partial_t =\Delta_{SL(d)}+ \frac{1}{4d}
\left(g_{ij} \frac{\partial}{\partial \tilde g_{ij}} \right)^2
\end{equation}
Together with \eqref{gllapa},
this yields the result \eqref{sllap} for the $SL(d)$
Laplacian.

\subsection{Laplacian on  the
$[SO(d)\times SO(d)]\backslash SO(d,d,\Real)$ symmetric space}
Next, we turn to the Laplacian on the  symmetric space
$[SO(d)\times SO(d)]\backslash SO(d,d)$ of dimension $d^2$. We can choose the
symmetric moduli matrix $M$ as in \eqref{modmat}, so that
the metric in \eqref{metric} reads
\begin{equation}
ds^2 = g^{ik} g^{jl} \left( dg_{ij} dg_{kl} + dB_{ij} dB_{kl}\right)
\end{equation}
This is a fibration on the coset $SO(d) \backslash Gl(d)$, so we
only need to compute the Laplacian on the fiber. We order the
indices as
\begin{equation}
ds^2_B = 2\sum_{i<j;k<l} \left( g^{ik} g^{jl} - g^{il} g^{jk}
\right) dB_{ij} dB_{kl}
\end{equation}
The determinant of the metric on the fiber is $\gamma_B=1/(\det
g)^{d-1}$, up to an irrelevant numerical factor. Using
\eqref{gldet}, the volume form on the total manifold is therefore
$\sqrt{\gamma}=(\det g)^{-d}$. The inverse metric reads
\begin{equation}
ds^2_{B,\rm inv} = \frac{1}{2}\sum_{i<j;k<l} \left( g_{ik} g_{jl}
- g_{il} g_{jk} \right) dB^{ij} dB^{kl}
\end{equation}
so that the Laplacian on the fiber is given by
\begin{equation}
\Delta_B = \frac{1}{2}\sum_{i<j;k<l} \left( g_{ik} g_{jl} - g_{il}
g_{jk} \right) \frac{\partial}{\partial B_{ij}}
\frac{\partial}{\partial B_{kl}}
\end{equation}
where we let $\partial B_{ij} /\partial
B_{kl} =\delta_i^k \delta_j^l  - \delta_i^l \delta_j^k$.
Putting this together with the Laplacian \eqref{gllap2}
on the base (with the appropriate volume element), we find
\begin{equation}
\Delta_{SO(d,d)}= \sum_{i\leq j;k\leq l} g_{ik} g_{jl}
\frac{\partial}{\partial g_{ij}} \frac{\partial}{\partial g_{kl}}
+ \sum_{i\leq j} g_{ij} \frac{\partial}{\partial g_{ij}} +
\frac{1}{4}\sum_{ijkl} g_{ik} g_{jl} \frac{\partial}{\partial
B_{ij}} \frac{\partial}{\partial B_{kl}}
\end{equation}
where the sum in the last term runs over unconstrained indices. An
alternative form using the redefined metric \eqref{gtil} is
\begin{equation}
\label{col} \Delta_{SO(d,d)} =
  \frac{1}{4} g_{ik} g_{jl}
 \left [ \frac{\pa}{ \pa \tilde{g}_{ij}} \frac{\pa}{ \pa \tilde{g}_{kl} }
 +  \frac{\pa}{ \pa B_{ij}} \frac{\pa}{ \pa B_{kl} }\right]
+   \frac{1}{2}  g_{ij} \frac{\pa}{ \pa \tilde{g}_{ij} }
\end{equation}
which is the one given in \eqref{solap}. The $SO(d,d+k)$ Laplacian
\eqref{asymlap} can be computed using similar techniques, but we
will not give the details of this computation here.

\subsection{Laplacian on the $U(g)\backslash Sp(g,\Real)$ symmetric space
\label{lape} }
Next, we derive the Laplacian on the $U(g)\backslash Sp(g)$
symmetric space, relevant for the genus $g$ amplitude in \eqref{gloop}.
Using for $M$ the moduli matrix \eqref{mosp}, the metric in \eqref{metric}
takes the form
\begin{equation}
ds^2=\tau_2^{AC} \tau_2^{BD} ( d\tau_{1AB}d\tau_{1CD}
+ d\tau_{2AB}d\tau_{2CD} )\ ,
\end{equation}
This is again  a fibration on the coset $SO(g)\backslash Gl(g)$, so
again we only need to compute the Laplacian on the fiber.
The determinant of the metric on the fiber is $\gamma|_{\tau_1}
=1/(\det \tau_2)^{g+1}$,
up to an (irrelevant) numerical factor, so that, using
\eqref{gldet} the volume form on the total
manifold is $\sqrt{\gamma}=(\det \tau_2)^{-(g+1)}$.
With the known result \eqref{gllap2} for the $Gl(d)$
Laplacian, we then obtain
\begin{multline}
\Delta_{Sp(g)}=
\sum_{A\leq B;C\leq D} \frac{\partial}{\partial \tau_{2AB}}
\tau_{2AC} \tau_{2BD} \frac{\partial}{\partial \tau_{2CD}}\\
+\tau_{2AC} \tau_{2BD}
\sum_{A\leq B;C\leq D} \frac{\partial}{\partial \tau_{1AB}}
\frac{\partial}{\partial \tau_{1CD}}
- (g+1) \sum_{A\leq B} \tau_{2AB}
\frac{\partial}{\partial \tau_{2AB}}
\end{multline}
Diagonally rescaling $\tau_1$ and $\tau_2$ as before gives the
more compact and covariant expression
\begin{equation}
\Delta_{Sp(g)}= \frac{1}{4}
 \tau_{2AC} \tau_{2BD}
\left( \frac{\partial}{\partial \tau_{1AB}}
\frac{\partial}{\partial \tau_{1CD}}
+\frac{\partial}{\partial \tau_{2AB}}
\frac{\partial}{\partial \tau_{2CD}} \right)
\end{equation}
which is the form given in \eqref{splap} and reduces to half the
usual Laplacian on the Poincar{\'e} upper half-plane for $g=1$.

\subsection{Laplacian on the $K\backslash \exc (\Real)$ symmetric
  space \label{slaped}}

We finally give here also the Laplacian on the the scalar manifold
$K\backslash \exc (\Real)$ of eleven-dimensional supergravity on
$T^{d+1}$ (equivalently type IIA string theory on $T^d$). In this
case, the scalars are given by the metric $g_{IJ}$, $I=1 \ldots
d+1$, a three-form $\C_{IJK}$ and its dual $\E_{6}$ (and for
$D=11-(d+1)\leq 3$ an extra $\K_{1,8}$-form, which will not be
included below). For $d \leq 6$, the corresponding Laplacian is
given by
\begin{multline}
\label{laped}
\Delta_{\exc}=
 \frac{1}{4} g_{IK} g_{JL}
\frac{\partial}{\partial g_{IJ}} \frac{\partial}{\partial g_{KL}}
+ \frac{(d+7)(d-4)}{4(d-8)}   g_{IJ} \frac{\partial}{\partial
g_{IJ}}
 + \frac{1}{4(8-d)}
  \left( g_{IJ}
\frac{\partial}{\partial g_{IJ}} \right)^2 \\
+ \frac{1}{2\cdot 3! ~l_M^6 } g_{IK} g_{JL} g_{PQ} \left(
\frac{\partial}{\partial \C_{IJP}} -10~\C_{RST}
\frac{\partial}{\partial \E_{RSTIJP}} \right) \left(
\frac{\partial}{\partial \C_{KLQ}} -10~\C_{UVW}
\frac{\partial}{\partial \E_{UVWKLQ}}\right) \\
+ \frac{1}{2 \cdot 6! ~l_M^{12} } g_{IK} g_{JL} g_{PQ} g_{RU} g_{SV} g_{TW}
\frac{\partial}{\partial \E_{IJPRST}}
\frac{\partial}{\partial \E_{KLQUVW}}
\end{multline}
The eigenvalues \eqref{psmeig} of the Eisenstein series of the
particle, string and membrane multiplet can be checked explicitly
from this form using the mass formulae of these multiplets and the
techniques employed in Appendix \ref{eig}. To this end it is
important to express the 11D Planck length $l_M$, which is not
invariant under the U-duality group $\exc (\Zint)$, in terms of
the invariant Planck length $l_P$ using the relation
$V_{d+1}/l_M^9= l_P^{d-8}$.

Note also that, since for $d\leq 4$ the U-duality groups $\exc
(\Zint)$ are of the $Sl$ and $SO$ type the Laplacian above should
reduce to the corresponding forms by appropriate redefinition of
the scalars. For $d=5,6$, with U-duality group $E_{6}$, $E_7$ the
above Laplacian is not contained in the previous results.

It is useful to determine the T-duality decomposition of the
Laplacian \eqref{laped}. For that purpose, we compute
the kinetic terms of the scalars in the Kaluza--Klein reduction
of ten-dimensional type IIA theory. Going to the Einstein
frame $g\to e ^{4\phi/(8-d)} g$, where $e ^{\phi}=g_s/\sqrt{V_d}$
is the invariant dilaton, we find
\begin{multline}
\label{iiared}
S=\int d^{10-d} x \sqrt{-g} \left[ R + \frac{4}{8-d}
\partial\phi\partial\phi -\frac{1}{4} \partial g\partial g^{-1}
+\frac{1}{4} \partial B g^{-1}\partial B g^{-1}+\right.\\
+\left. \frac{e^{2\phi}}{2}~
\partial {\mathcal R} \cdot M({\spi}) \cdot \partial {\mathcal R}
+\dots\right]
\end{multline}
Here, $\R$ denote the Ramond scalars transforming in the
spinor representation of $SO(d,d)$, and the dots stand for
extra scalars which originate from dualizing the
Kaluza--Klein one-form, Neveu-Schwarz two-form or Ramond
forms in $d\ge 5$. From the property
\begin{equation}
\sum_{k=even} k {d \choose k}=\sum_{k=odd} k {d\choose k} = d~2^{d-2}
\ ,
\end{equation}
it follows that the mass matrix $M({\spi})$, like $M(\spb)$,
has unit determinant.
The volume element is thus given by $\sqrt{\gamma}=e ^{2^{d-1}\phi}$
(for $d <  5$ and in fact also $d=5$), and the Laplacian on
the symmetric space $K\backslash \exc(\Real)$ then reads, in
variables appropriate for T-duality,
\begin{equation}
\label{lapedt}
\Delta_{E_{d+1(d+1)}}=\frac{8-d}{16} \left( \partial_\phi ^2+
2^{d-1}\partial_\phi \right) + \Delta_{SO(d,d)} +
\frac{e^{-2\phi}}{2}
\partial_\R \cdot M^{-1}({\spi})\cdot \partial_\R+\dots
\end{equation}
From this we can for example check that the Einstein-frame
tree-level $R^4$ term $e ^{12\phi/(d-8)}$, or the one-loop
term $e ^{2(d-2)\phi/(d-8)} \eis{SO(d,d,\Zint)}{\spi,\spb}{s=1}$
are eigenmodes of the U-duality invariant Laplacian as
required by the conjecture \eqref{npeig}.

\subsection{Decompactification of the Laplacians \label{dec} }

We conclude by giving the decompactification formulae for the
$Gl(d)$ and $SO(d,d)$ Laplacians. These are relevant for the study
of the decompactifcation properties of the corresponding
Eisenstein series.

We will consider only the $SO(d,d)$ case, since the resulting
formulae for $Gl(d)$ and $SL(d)$ can easily be obtained from this
case. For the metric we take the $U(1)$-fibered form
\begin{equation}
 d x^i g_{ij} \ d x^j= R^2 (dx^1 + A_a d x ^a)^2 + d x^a \hg_{ab} d x^b
\end{equation}
where $a = 2, \ldots d$ and the original metric is $g_{ij}$. We
also define
\begin{equation}
B_{1a} = B_{a} \sp B_{ab}= \hB_{ab}+ \frac{1}{2} [ A_a B_b - A_b
B_a ]
\end{equation}
In terms of these variables, T-duality takes the simple form
\begin{equation}
\label{tdua} R\leftrightarrow \frac{1}{R}\ ,\quad e
^{\phi}\leftrightarrow \frac{e ^{\phi}}{R}\ ,\quad
A_a\leftrightarrow B_a\ ,\quad (\hg_{ab},\hB_{ab}) \ \mbox{inv.}
\end{equation}
For the purpose of dimensional reduction it is, however, more
convenient to introduce a modified $\tB_{ab}$ field invariant
under gauge transformations of $A_a$ (but not under shifts of
$B_a$):
\begin{equation}
B_{ab} = \tB_{ab} + A_a B_b - B_a A_b
\end{equation}
In the expressions below, we also use the diagonally
rescaled metric \eqref{gtil} for $\hg$ whenever it appears in derivatives.
The Jacobian for
the change of variables from $(g_{ij},B_{ij})$ to
$(R,A_a,\hg_{ab},B_a,\tB_{ab})$ is given by
\begin{subequations}
\label{jacob}
\begin{equation} \frac{\pa}{ \pa g_{11} } =
\frac{1}{2R} \frac{\pa}{ \pa R} -\frac{A_a}{R^2} \frac{\pa}{ \pa
A_a} + \frac{1}{2} A_a A_b \frac{\pa}{ \pa \hg_{ab} } +\frac{A_a
B_b}{R^2} \frac{\pa}{\pa \tB_{ab}}
\end{equation}
\begin{equation}
\frac{\pa}{ \pa g_{1a} } = \frac{1}{R^2} \frac{\pa}{ \pa A_a} -  A_b
\frac{\pa}{ \pa \tilde{\hg}_{ab} }-\frac{B_b}{R^2} \frac{\pa}{\pa
\tB_{ab}} \sp
\frac{\pa}{ \pa g_{ab} } = \frac{\pa}{ \pa \hg_{ab} }
\end{equation}
\begin{equation}
\frac{\pa}{\pa B_{1a}} = \frac{\pa}{\pa B_{a}} + A_b
\frac{\pa}{\pa \tB_{ab}} \sp \frac{\pa}{\pa B_{ab}} =
 \frac{\pa}{\pa \hB_{ab}}
\end{equation}
\end{subequations}
The Jacobian relevant for the $Gl(d)$ Laplacian is simply obtained
by ignoring the terms involving $B$. Then, we find for the $Gl(d)$
Laplacian the decomposed result \eqref{gllapd},
while for $SO(d,d)$ we have after some algebra
\begin{multline}
\label{dsored} \Delta_{SO(d+1,d+1)} =\Delta_{SO(d,d)} - \frac{1}{2}
 \hg_{ab} \frac{\partial}{\partial \hg_{ab}}+
\frac{1}{4} \left( R\frac{\partial}{\partial R} \right)^2 +
\frac{\hg_{ab}}{2R^2} \frac{\partial}{\partial
A_a}\frac{\partial}{\partial A_b}\\
+ \frac{R^2 \hg_{ab}}{2} \frac{\partial}{\partial
B_a}\frac{\partial}{\partial B_b} -\frac{1}{R^2} \hg_{ab} B_c
\frac{\partial}{\partial A_a}\frac{\partial}{\partial \tB_{bc}}
+\frac{1}{2R^2}\hg_{ab} B_c B_d \frac{\partial}{\partial \tB_{ab}}
\frac{\partial}{\partial \tB_{cd}}
\end{multline}

We also note that the corresponding Jacobian for the change to the
$(R,A_a,\hg_{ab},B_a,\hB_{ab})$ variables can be obtained from the
one in \eqref{jacob} by substituting $\tB \rightarrow 2\hB$ except
for the last equation. For completeness, we also give the
decomposed $SO(d,d)$ Laplacian in these variables
\begin{multline}
\Delta_{SO(d+1,d+1)} =\Delta_{SO(d,d)} - \frac{1}{2} \hg_{ab}
\frac{\partial}{\partial \hg_{ab}}+ \frac{1}{4} \left(
R\frac{\partial}{\partial R} \right)^2 + \frac{\hg_{ab}}{2R^2}
\frac{\partial}{\partial A_a}\frac{\partial}{\partial
A_b} + \frac{R^2 \hg_{ab}}{2} \frac{\partial}{\partial
B_a}\frac{\partial}{\partial B_b} \\
+\frac{1}{8} \hg_{ac} \left(R^2 A_b A_d + \frac{B_b B_d}{R^2}
\right) \frac{\partial}{\partial \hB_{ab}}
\frac{\partial}{\partial \hB_{cd}}   - \frac{1}{2}
\hg_{ab}\left(R^2 A_c\frac{\partial}{\partial B_{a}} +
\frac{B_c}{R^2}\frac{\partial}{\partial A_{a}} \right)
 \frac{\partial}{\partial
\hB_{bc}}
\end{multline}
which manifestly exhibits the T-duality symmetry \eqref{tdua}.

\section{Eigenmodes and eigenvalues of the Laplacians \label{eig} }

In this appendix we give some details on the explicit computation
of the eigenvalues under the Laplacian and the non-invariant
differential operator \eqref{box} of the various Eisenstein series
and modular integral considered in the main text. These computations
are most easily done using the integral representation
\begin{equation}
\label{intrep2} \left[\M^2\right]^{-s}=\frac{\pi ^s}{\Gamma(s)}
\int_{0}^{\infty} \frac{dt}{t^{1+s}}  \exp \left( -\frac{\pi} {t}
\M^2 \right)
\end{equation}
of the generic term in the Eisenstein series.
The result of differentiation can be integrated by parts using
\begin{equation}
\label{ide} \int_0^{\infty} \frac{dt}{t^{1+s}} \left[
\alpha \frac{C^2}{t^2} + \beta \frac{C}{t} \right] e^{ - C/t}
= s( \alpha s+ \alpha + \beta ) \int_0^{\infty}
\frac{dt}{t^{1+s}}  e^{-C/t}
\end{equation}

\subsection{$SL(d,\Zint)$ Eisenstein series in the fundamental representation}
We start with the fundamental representation of $SL(d,\Zint)$,
for which the mass matrix reads
$\M^2 (\irrep{d}) = m^t g m = m^i g_{ij} m^j$,
and obeys the identities
\begin{equation}
\frac{\pa \M^2 (\irrep{d})}{\pa g_{ij} } =  2 m^i m^j
\ ,\qquad \frac{\pa^2 \M^2 (\irrep{d})}{\pa g_{ij} \pa g_{kl} }
= 0
\end{equation}
so that using the Laplacian \eqref{gllap}, we obtain, setting $t'=t/\pi$,
\begin{multline}
 e^{ \M^2 (\irrep{d}) /t'}
\Delta_{Gl(d)} e^{-\M^2 (\irrep{d}) /t'}  =
 \frac{1}{t'^2}  g_{ik} g_{jl} (    m^i m^j ) (  m^k  m^l)
-  \frac{d+1}{4t'} g_{ij}  2 m^i m^j
\\
= \frac{1}{t'^2} \left[\M^2 (\irrep{d}) \right]^2 - \frac{d+1}{2t'}
\M^2 (\irrep{d})
\end{multline}
Then, using the identity \eqref{ide}
we immediately find the eigenvalue $s(s +1 - \frac{d+1}{2} )$ as given
in \eqref{glfun}. The corresponding eigenvalue under the $SL(d)$ Laplacian
follows by subtracting the $(t\pa/\pa t)^2/d$ contribution in
\eqref{glsl},
\begin{equation}
\frac{1}{4d}
 e^{\M^2 (\irrep{d}) /t'}
\left(g_{ij} \frac{\partial}{\partial g_{ij}} \right)^2
 e^{-\M^2 (\irrep{d}) /t'}  =
\frac{1}{d} \left( \frac{1}{t'^2} \left[\M^2 (\irrep{d}) \right]^2 -
\frac{1}{t'}
\M^2 (\irrep{d}) \right)
\end{equation}
so that the eigenvalue is  $s(s +1 - (d+1)/2 ) -
s^2/d$ as given in \eqref{esllap}.

\subsection{$SO(d,d,\Zint)$ Eisenstein series in the vector representation}
For the case of the vector representation of $SO(d,d)$, the mass matrix
now reads
\begin{equation}
\M^2 (\vect) = m^t M (\vect) m = \tilde m_i g^{ij} \tilde m_j  +
n^i g_{ij} n^j \ ,\qquad \tilde m_i = m_i + B_{ij} n^j
\end{equation}
and satisfies
\begin{equation}
\frac{\pa \M^2 (\vect)}{\pa g_{ij}} =
2 \left[- \tilde m^i \tilde m^j + n^i  n^j \right]
\sp
\frac{\pa \M^2 (\vect)}{\pa B_{ij}} =
2 \left[ \tilde m^i  n^j -  \tilde m^j   n^i  \right]
\end{equation}
where $\tilde m^i = g^{ij} \tilde m_j$. To compute the action of
the Laplacian \eqref{solap} and of the operator $\square_d$ in
\eqref{box} we need the quantities
\begin{subequations}
\begin{eqnarray}
D_0 &=&  \left( \frac{1}{2} g_{ij} \frac{ \pa \M^2(\vect)}{\pa g_{ij} }
\right)^2 = (\tilde m^2 )^2 + ( n^2)^2  - 2 \tilde m^2   n^2\\
D_1 &=& \frac{1}{4} g_{ik} g_{jl} \frac{ \pa \M^2 (\vect) }{\pa
g_{ij} } \frac{ \pa \M^2 (\vect)}{\pa g_{kl} } = (\tilde m^2 )^2 +
( n^2)^2  - 2 (\tilde m  n)^2\\
D_2 &=& \frac{1}{4} g_{ik} g_{jl} \frac{ \pa \M^2(\vect)}{\pa B_{ij}
} \frac{ \pa \M^2(\vect)}{ \pa B_{kl} } = 2[ \tilde m^2  n^2 -
(\tilde m  n)^2 ]
\end{eqnarray}
\end{subequations}
as well as
\begin{subequations}
\begin{equation}
C_1 = \frac{1}{4} g_{ik} g_{jl} \frac{ \pa^2 \M^2
(\vect)}{\pa g_{ij}\pa g_{kl}} = (d+1) \tilde m^2 \sp
C_2 = \frac{1}{4}
g_{ik} g_{jl}  \frac{\pa^2 \M^2 (\vect)}{
\pa B_{ij} \pa B_{kl}} = (d-1)  n^2
\end{equation}
\begin{equation}
C_0 = \frac{1}{4} g_{ij} \frac{\pa}{ \pa g_{ij}} g_{kl}\frac{ \pa \M^2
(\vect)}{\pa g_{kl}} = \tilde m^2 +  n^2 \sp C = \frac{1}{2}
g_{ij}\frac{ \pa \M^2 (\vect)}{ \pa g_{ij}} = -\tilde m^2 +  n^2
\end{equation}
\end{subequations}
Using these data it is then easy to compute
\begin{subequations}
\begin{align}
e^{\M^2 (\vect) /t'}
\Delta_{SO(d,d)} e^{-\M^2 (\vect) /t}  =&
 \frac{1}{t'^2} \left[D_1 + D_2\right]
-  \frac{1}{t'} \left[ C_1 + C_2 + C\right]
\\ =&\frac{1}{t'^2} \left[(\M^2(\vect))^2 - 4 (mn)^2\right] -
\frac{d}{t'} \M^2 (\vect) \\ 
e^{\M^2 (\vect) /t'} \square_d \exp
e^{-\M^2 (\vect) /t'}  =&
 \frac{1}{t'^2} \left[D_1 -\frac{1}{2} D_0 \right]
-  \frac{1}{t'} \left[C_1 -\frac{1}{2} C_0 + \frac{1}{2}(d+1)C \right]
\\
=& \frac{1}{2t'^2} \left[(\M^2(\vect))^2 - 4 (mn)^2\right] -
\frac{d}{2t'} \M^2 (\vect)
\end{align}
\end{subequations}
The eigenvalues $s(s-d+1)$ and
$\frac{s}{2}(s-d+1)$ of the vector Eisenstein series under the
Laplacian $\Delta_{SO(d,d)}$ and the non-invariant operator
$\square_d$ follow by using the identity \eqref{ide},
provided the half-BPS constraint
$\tilde m n = mn=0$ is satisfied. These are the values quoted in
\eqref{soeigval} and \eqref{sqeigval} respectively.

\subsection{$SO(d,d,\Zint)$ Eisenstein series in the spinor representations}
The direct computation of the eigenvalues of the (conjugate)
spinor Eisenstein under the $SO(d,d)$ Laplacian is more involved
and will not be given here.  The general results in
\eqref{soeigval} have been checked directly for $d\leq 4$, showing
also in this case explicitly the importance of imposing the
half-BPS constraints \eqref{13con} and \eqref{22con} that occur
for $d=4$.

Finally, we turn to the action of the $\square_d$ operator on the
(conjugate) spinor  representation \eqref{dstring},
\eqref{dparticle} of $SO(d,d)$, with mass
\begin{equation}
\label{mspi} \M^2 (\spi) = \frac{1}{V_d} \sum_{p=\mbox{odd}}  \frac{(\tilde
m^{[p]})^2}{p!} \sp
\M^2 (\spb) = \frac{1}{V_d} \sum_{p=\mbox{even}} \frac{(\tilde
m^{[p]})^2}{p!}
\end{equation}
where $\tilde m^{[p]} = m^{[p]} + B_2 m^{[p-2]} + B_2^2 m^{[p-4]} + \ldots$
are the dressed charges and $(\tilde m^{[p]})^2$ denotes the invariant square
obtained with $p$ powers of the metric. We need the derivative
\begin{equation}
\label{mspid}
\frac{\pa \M^2 (\spi)}{\pa g_{ij}}
 = \frac{1}{V_d} \sum_{p} \frac{2 p
[(\tilde m^{[p]})^2]^{ij} -  (\tilde m^{[p]})^2 g^{ij} }{p!}
\end{equation}
where  $[(\tilde m^{[p]})^2]^{ij}$ denotes the invariant square
with one power of the metric taken out. The direct computation of
the full $\square_d$ along the same lines as the cases treated
above is rather intricate. We therefore employ a method that uses
the underlying group theory and the realization that $\square_d$
contains the $SL(d)$ Laplacian, as well as the structural form
\eqref{dsconstr} of the constraints.

We first note that each term in \eqref{mspi} represents a totally
antisymmetric tensor of $SL(d)$ with $p$ indices.  For an
antisymmetric $p$-tensor of $SL(d)$, the Casimir of the $r$th
symmetric power is given by
\begin{equation}
Q( \irrep{[p]^{\otimes_s r}}) =\frac{rp (d-p)(r+d)}{d}
\end{equation}
so that according to the general formula \eqref{eisgen} the action of
the $SL(d)$ Laplacian is
\begin{equation}
\label{casanti}
\Delta_{SL(d)}[ (m^{[p]})^2 ]^{-s} = \frac{p(d-p)s(2s-d)}{2d} [(m^{[p]})^2]^{-s}\sp
\end{equation}
Using the identity \eqref{ide}, we therefore have, up to cross terms
which we neglect for the moment,
\begin{multline}
\label{boxs1}
 e^{\M^2 (\spi)/t'}
\Delta_{SL(d)} e^{-\M^2 (\spi)/t'}   =\\
\frac{1}{t'^2} \left[ \sum_{p}  \frac{p(d-p)}{d}
\left(\frac{(m^{[p]})^2}{V_d~p!} \right)^2 + {\rm cross}\right]
-\frac{1}{t'} \left[ \sum_p \frac{p(d-p)(d+2)}{2d} \frac{(m^{[p]})^2}{V_d~p!} \right]
\end{multline}
where we emphasize that this  result is only valid when enforcing
the quadratic constraints on the charges. We also need
\begin{equation}
\label{boxs2}
e^{\M^2 (\spi)/t'}
\left(  \frac{1}{2} g_{ij} \frac{\pa}{ \pa g_{ij} }\right)^2
e^{-\M^2 (\spi)/t'}  =
 \frac{1}{t'^2} \left[ \sum_p \left(p-\frac{d}{2}\right)
\frac{(m^{[p]})^2}{V_d~p!} \right]^2 -\frac{1}{t'} \left[ \sum_p
\left(p-\frac{d}{2}\right)^2 \frac{(m^{[p]})^2}{V_d~p!} \right]
\end{equation}
obtained by direct calculation using \eqref{mspid}. Using the form
of $\square_d$ in \eqref{box} we obtain from the two expressions
in \eqref{boxs1} and \eqref{boxs2} that
\begin{multline}
\label{boxs}
 e^{\M^2 (\spi)/t'}
\square_d\
 e^{-\M^2 (\spi)/t'}
=\frac{1}{t'^2} \left[
\sum_p
\left(\frac{p(d-p)}{2}-\frac{d(d-2)}{8}\right)
\left(\frac{(m^{[p]})^2}{V_d~p!}\right)^2 +
  {\rm cross} \right]
\\
+\frac{1}{t'} \left[ \sum_p
\left(-p(d-p)+\frac{d(d-2)}{8}\right)
\frac{(m^{[p]})^2}{V_d~p!} \right]
\end{multline}
Using \eqref{ide} we then deduce that
\begin{multline}
\label{boxspi} \frac{\square_d (\M^2(\spi))^{-s}}{(\M^2 (\spi)
)^{-s-2}}
 = s \left\{
(s+1)\left[ \sum_p
\left(\frac{p(d-p)}{2}-\frac{d(d-2)}{8}\right)
\left(\frac{(m^{[p]})^2}{V_d~p!}\right)^2 +{\rm cross}  \right]
\right.\\
\left.
+ \M^2 (\spi) \sum_p
\left(-p(d-p)+\frac{d(d-2)}{8}\right)
\frac{(m^{[p]})^2}{V_d~p!}
\right\}
\end{multline}
Requiring the righthand side to be proportional to (the diagonal
terms in) $(\M^2 (\spi))^2$ then shows us that (for generic value
of $d$) this is only possibly when $s=1$, in which case we find
that
\begin{equation}
\square_d (\M^2 (\spi) )^{-s}=\frac{d(2-d)}{8} ( \M^2 (\spi)
)^{-s}\ ,\qquad s=1
\end{equation}
so that the $s=1$ spinor and conjugate spinor Eisenstein series
are eigenmodes of $\square_d$ as recorded in \eqref{bspieig},
\eqref{bspbeig}.

A special feature  arises for the spinor representation of
$SO(4,4)$, in which case we have $p(d-p) =p(4-p) = 3$ for both the
relevant values $p=1$ and 3 so that the terms in \eqref{boxspi}
are proportional to $(\M^2 (\spi))^2$
 for all values of $s$. As a result we find
that the spinor Eisenstein series for $SO(4,4)$ is an eigenmode of
$\square_d$ with eigenvalue $s(s-3)/2$ as noted in \eqref{so44}
\footnote{In a similar way one can see that the spinor and
conjugate spinor Eisenstein series of $SO(2,2)$ are eigenvalues
for all $s$, but that was expected since in that case $\square_d$
reduces to the $SL(2)$ Laplacian.}. This is not the case for the
conjugate spinor representation of $SO(4,4)$.

Finally, we wish to point out some further checks on the cross
terms that have been neglected so far. First, we have explicitly
checked the full result by direct computation for the spinor
representation in the cases $d\leq 4$. In particular, for $d=4$
one finds, as expected that the constraint $m^{[1]} \wedge m^{[3]}=0$ of
\eqref{13con} is crucial for the eigenvalue condition. More
generally, using the metric on weight space
$g_{\irrep{[p][q]}}=p(d-q)/d$
with $\irrep{[p]}\leq \irrep{[q]}$
two totally antisymmetric $SL(d)$ representations, we
know from the group theory arguments in \eqref{delexp} that the cross
terms in \eqref{boxs1} can be incorporated by replacing the
$1/t'^2$ term by
\begin{equation}
\frac{1}{t'^2} \left[ \sum_{p\leq q} (2-\delta_{pq}) \frac{p(d-q)}{d}
\left(\frac{(m^{[p]})^2}{V_d~p!}\right)\left(\frac{(m^{[q]})^2}{V_d~q!}\right)
\right]
\end{equation}
Together with the directly computed cross terms in \eqref{boxs2}, this
changes the $1/t'^2$ term in \eqref{boxs} to
\begin{equation}
\frac{1}{t'^2} \sum_{p\le q} (2-\delta_{pq})
\left(\frac{d(p+q)+2(p-q)-2pq}{4}-\frac{d(d-2)}{8}\right)
\left(\frac{(m^{[p]})^2}{V_d~p!}\right)\left(\frac{(m^{[q]})^2}{V_d~q!}\right)
\end{equation}
which will produce an analogous correction to \eqref{boxspi}.
For $s=1$ we then see that, taking into account the cross terms from
the second term in \eqref{boxspi}, the $(p,q)$-dependent part is
given by
\begin{equation}
\label{cross}
\left(d(p+q)+2(p-q)-2pq\right)- \left( p (d-p)+q(d-q) \right)
=(p-q)(p-q+2)
\end{equation}
which we see vanishes (besides the diagonal terms $p=q$)  for
the cross terms $q-p=2$.

If $q-p>2$, there are non-trivial effects from the constraints. A
simple way to see them is to consider the two-form in the $d=4$
conjugate spinor. The $1/t^2$ contribution to $\Delta_{Sl}$
includes a term $(m^{[2]})^4$, where the contraction is the
non-factorized one. By the Cayley--Hamilton theorem for $4\times 4$
antisymmetric matrices $A$,
\begin{equation}
A^4 - \frac{1}{2} (\tr A^2) A^2 + (\pf A)^2 1 = 0
\end{equation}
we see that this is $[(m^{[2]})^2]^2/2$ up to a $(m^{[2]}\wedge m^{[2]})^2$ term,
which by the half-BPS condition \eqref{22con} is equivalent to an
extra cross term $(m~m^{[4]})^2$ that was not taken into account
previously and will cancel the deficit seen in \eqref{cross}.

\section{Large volume expansions of Eisenstein series \label{lve}}

Here, we derive the results \eqref{largevolV}, \eqref{largevol} by
considering the large volume expansions of the vector, spinor and
conjugate spinor Eisenstein series of $SO(d,d)$. In the
computations below we shall repeatedly use the Poisson resummation
formula
\begin{equation}
\label{pois}
\sum_{m} e^{-\pi (m +a)^t A (m+a) + 2 \pi i m b }
= \frac{1}{\sqrt{\det A}} \sum_{\tilde m} e^{-\pi
(\tilde m +b)^t A^{-1} (\tilde m+b)
 - 2 \pi i (\tilde m +b) a }
\end{equation}
Note that an insertion of $m$ on the lefthand side translates into
an insertion of $-a+i A^{-1}(\tilde m+b)$ on the righthand side.
We also recall the integral representation of the Bessel function
\begin{equation}
\label{bessel}
\int_0^{\infty} \frac{dx}{x^{1+s}} e^{-b/x-cx}
=2 \left|\frac{c}{b }\right|^{s/2} K_{s}(2\sqrt{|bc|})
\end{equation}
It is an even function in $s$, and admits the asymptotic expansion
at large $x$
\begin{equation}
\label{asbessel}
K_{s}(x)= \sqrt{\frac{\pi}{2x}} e^{-x} \left(1
+ \sum_{k=1}^{\infty} \frac{1}{(2x)^k}
\frac{\Gamma\left(s+k+\frac{1}{2}\right)}
{k!\Gamma\left(s-k+\frac{1}{2}\right)}
\right)\ .
\end{equation}
The expansion truncates when $s$ is half-integer,
and in particular, for $s=1/2$ the saddle point approximation is exact:
\begin{equation}
\label{bessel1}
K_{1/2}(x)=\sqrt{\frac{\pi}{2x}} e^{-x}
\end{equation}
We also recall some useful facts about the
Riemann Zeta and Gamma functions
\begin{subequations}
\begin{gather}
\zeta(s) = \sum_{m=1}^{\infty} \frac{1}{m^s}
=\frac{\pi^{s/2} \Gamma(1-s/2)}{\pi^{(1-s)/2} \Gamma(s/2)}
\zeta(1-s)\\
\label{zetav}
 \zeta(-1) = -\frac{1}{12} \ ,\quad
\zeta (0) = -\frac{1}{2} \ ,\quad \zeta (2) = \frac{\pi^2}{6}\ ,\\
\Gamma (s) = \int_0^{\infty} \frac{dt}{t^{1+s}} e ^{-1/t} \sp
 \Gamma (s+1) = s \Gamma(s) \ ,\quad \Gamma (1) = 1 \ ,\quad
 \Gamma(1/2)=\sqrt{\pi}\ .
\end{gather}
\end{subequations}
It is also useful to recall that
$\zeta(s)$ has a simple pole at $s=1$, simple zeros
at $s=-2,-3,\dots$ whereas $\Gamma(s)$ has simple poles at $s=0,-1,-2,\dots$:
\begin{equation}
\zeta(1+\epsilon)=\frac{1}{\epsilon}+\gamma+{\cal O}(\epsilon)
\sp
\Gamma(\epsilon)=\frac{1}{\epsilon} -\gamma+{\cal O}(\epsilon)
\end{equation}
where $\gamma=0.577215...$ is the Euler constant.

\subsection{$SO(d,d,\Zint)$ vector Eisenstein series}
We first consider the large volume expansion of the Eisenstein series in
the vector representation of $SO(d,d)$, for which we use the integral
representation
\begin{equation}
\eis{SO(d,d,\Zint)}{s}{\vect}=\frac{\pi ^s}{\Gamma(s)}
\int_{0}^{\infty} \frac{dt}{t^{1+s}}\int_0^1 d\theta
\hat{\sum_{m_i,n^i}}  \exp \left( -\frac{\pi}{t}
(m_i+B_{ij}n^j)^2 -\frac{\pi}{ t} (n^i)^2  + 2\pi i \theta m_i n^i
\right)
\end{equation}
Here the integration over $\theta$ incorporates the constraint
$m_i n^i =0$ and the squares denote the invariant contraction with
the metric or inverse metric depending on the position of the
indices. We first extract the $n^i=0$ piece, and in the remaining
part Poisson resum on the integers $m_i$ which are now
unconstrained. Then $\eis{SO(d,d,\Zint)}{\vect}{s}=J_1 (\vect) +
J_2(\vect)$ with
\begin{eqnarray}
\label{jv1}
J_1(\vect) &=&  \hat{\sum_{m_i}}  \left[ \frac{1}
{m_i g^{ij}m_j}\right]^{s}
= \eis{SL(d,\Zint)}{\irrep{\bar{d}}}{s} =
V_d \frac{\pi^{s} \Gamma(\frac{d}{2}-s) }
{  \pi^{\frac{d}{2}-s} \Gamma(s) }
 \eis{SL(d,\Zint)}{d}{\frac{d}{2}-s} \\
\label{jv2} J_2(\vect) &=& \frac{V_d \pi ^s}{\Gamma(s)}
\int_{0}^{\infty} \frac{dt}{t^{1+s-\frac{d}{2}}}\int_0^1 d\theta
\sum_{m^i} \hat{\sum_{n^i}} \exp \left( -\pi t (m^i+\theta n^i)^2
-\frac{\pi}{t}  (n^i)^2  + 2\pi i B_{ij} m^i n^j \right) \nonumber
\end{eqnarray}
Here we have recognized the first term \eqref{jv1} as the Eisenstein series of
the antifundamental of $SL(d)$ and used the identity \eqref{slinv}
in the last step.
Continuing with the second term \eqref{jv2} we note that although
the integration over $\theta$ runs from 0 to 1 only, we can reabsorb
a shift $\theta\rightarrow\theta+1$ into a spectral flow
$m_i\rightarrow m_i+n_i$. We therefore extend
the integration range of $\theta$ to $N \rightarrow \infty $ but sum on
$m_i$ modulo $n_i$ only. Then, after performing the Gaussian integration
on $\theta$, the second term becomes
\begin{equation}
\label{jv2p} J_2 (\vect) = \frac{V_d \pi ^s}{\Gamma(s)}
\int_{0}^{\infty} \frac{dt}{t^{1+s-\frac{d-1}{2}}}
\hat{\sum_{m^i \,\mbox{\scriptsize mod}\,  n^i} }
\frac{
\exp \left( { -\pi t
\frac{(m\cdot m)(n\cdot n)-(m\cdot n)^2}{n\cdot n}
-\frac{\pi}{t}   n\cdot n  + 2\pi i B_{ij} m^i n^j} \right) }
{\sqrt{ n^i g_{ij} n^j}}
\end{equation}
We now extract the terms for which $(m\cdot n)^2-(m\cdot
m)(n\cdot n)=0$. By Schwarz inequality, this is the case if
and only if $m_i=\lambda n_i$ for all $i$, and therefore
the phase factor in \eqref{jv2p} is irrelevant. For a given
vector $n$, the number of parallel vectors $m$ modulo the spectral
flow is $gcd(\{n_i\})$, so that we have
$J_2 (\vect) = J_{2a} (\vect) + J_{2b} (\vect )  $ with
\begin{multline}
\label{jv2a}
 J_{2a} (\vect) =
\frac{V_d \pi^{\frac{d-1}{2}} \Gamma(s-\frac{d-1}{2})}{\Gamma(s)}
\hat{\sum} \gcd(\{n^i\}) \left[ \frac{1}{n^i g_{ij} n^j} \right]^{s-\frac{d-2}{2}}
=\\
=\frac{V_d \pi^{\frac{d-1}{2}} \Gamma(s-\frac{d-1}{2})
\zeta(2s-d+1)}{\Gamma(s)\zeta(2s-d+2)}
\eis{SL(d,\Zint)}{d}{s-\frac{d}{2} +1 }
\end{multline}
Here we have split the integers $n^i$ into coprime $n^{'i}$'s and
greatest common divisor $r$, carried out the $r$-summation, and
rewritten  the coprime integers in terms of integers again at the
expense of yet another $r$ summation. Finally, for the remaining
non-degenerate terms $J_{2b} (\vect)$ in \eqref{jv2p} we can
perform the integral on $t$ using \eqref{bessel} so that
\begin{multline}
\label{jv2b}
J_{2b} (\vect) =\frac{4V_d \pi^{s}}{\Gamma(s)}
\hat{\sum_{m^i \,\mbox{\scriptsize mod}\,  n^i} } \frac{1}{\sqrt{n^i g_{ij} n^j}}
\left[\frac{(n\cdot n)^2}{|(m\cdot m)(n\cdot n)-(m\cdot n)^2|}\right]
^\frac{d-1-2s}{4}\\
K_{s-\frac{d-1}{2}} \left(2\pi
 \sqrt{| (m\cdot m)(n\cdot n)-(m\cdot n)^2|} \right)
e ^{2\pi i B_{ij} m^i n^j }
\end{multline}
As in the  non-degenerate orbit contribution of the one-loop integral, it is convenient
to decompose the set of integers $m^i,n^i$ with $m^{ij}\neq 0$ into $SL(2,\Zint)$
equivalence classes, and write 
\be
\begin{pmatrix} m^i \\ n^i \end{pmatrix} = \gamma \cdot 
\begin{pmatrix} \bar m^i \\ \bar n^i \end{pmatrix} \ ,\qquad
\gamma = \begin{pmatrix} a & b \\c & d \end{pmatrix} \in \Gamma
\ee
where $(\bar m^i,\bar n^i)$ is any coset representative in the 
equivalence classe labelled by the rank 2 matrix $d^{ij}=\frac12(
\bar m^i \bar n^j- \bar n^i \bar m^j)$.
The equivalence relation $m^i\equiv m^i+ n^i$ means that $\gamma$ must
run over $\Gamma_\infty\backslash \Gamma$ only. Thus, we can rewrite 
\begin{equation}
\begin{split}
\label{jv2b2}
J_{2b}  &=\frac{4V_d\, \pi^{s}}{\Gamma(s)}
\sum\limits_{\substack{d^{ij} \neq 0\\ \rank(d^{ij})=2}}
\mathcal{E}(d^{ij})\, 
\left[{(d^{ij})^2}\right]
^\frac{-d+1+2s}{4}\, 
K_{s-\frac{d-1}{2}} \left(2\pi
 \sqrt{(d^{ij})^2} \right)
e ^{2\pi i B_{ij} d^{ij} }\ ,
\end{split}
\end{equation}
where
\be
\mathcal{E} (d^{ij}) = 
\sum_{(c,d)=1} \left\|c \bar m + d \bar n\right\|^{d-2-2s} \ .
\ee
This is recognized as an Eisenstein series \eqref{slef} for $d=2$, evaluated at the Gram matrix
of $\bar m$ and $\bar n$,
\be
\mathcal{E}(d^{ij}) =
 \frac{1}{2 \zeta( 2s+2-d)}\, \eis{SL(2,\Zint)}{\irrep{2}}{ s+1-\frac{d}{2} }
 \left({\scriptsize  \begin{matrix} \bar m\cdot \bar m & 
\bar m\cdot \bar n\\
\bar m \cdot \bar n & \bar n\cdot \bar n \end{matrix}} \right) \ .
\ee
(by $SL(2,\Zint)$ invariance, the r.h.s. only depends on $d^{ij}$).
At the special value $s=\frac{d}{2}-1$, collecting these results and using \eqref{ana} 
we find that the Eisenstein series in the vector representation reduces to
\begin{multline}
\label{lvV}
\eis{SO(d,d,\Zint)}{\vect}{s=\frac{d}{2}-1}=
\frac{  \pi^{\frac{d}{2}-2}  }{\Gamma( \frac{d}{2} -1)}
\left[ V_d \eis{SL(d,\Zint)}{\irrep{d} }{s=1}  + \frac{\pi^2}{3} V_d +
\right. \\
+ \left. 2\pi V_d
\bar{\sum_{m^i,n^i} } \frac{
\exp\left(-2\pi \sqrt{|(m\cdot m)(n\cdot n)-(m\cdot n)^2|}
+ 2\pi i B_{ij} m^i n^j \right) }
{\sqrt{|(m\cdot m)(n\cdot n)-(m\cdot n)^2}|} \right]
\end{multline}
where the sum now runs over non-degenerate $SL(2,\Zint)$
orbits of vectors $(m,n)$.
Here, to simplify the second term \eqref{jv2a} we have used
\eqref{zetav} and $\eis{SL(d,\Zint)}{\irrep{d}}{s=0} = -1$ (see
\eqref{ana}). For the third term \eqref{lvV} we have used
\eqref{bessel1} to express the Bessel function $K_{1/2}$ as an
exponential. We have thus reproduced the announced result
 \eqref{largevolV}.

\subsection{$SO(3,3,\Zint)$ spinor and conjugate spinor Eisenstein series}
We start with the spinor representation of $SO(3,3)$ with Eisenstein
series with integral representation,
\begin{equation}
\eis{SO(3,3,\Zint)}{\spi}{s}
=\frac{\pi^s}{\Gamma(s)}
\int_{0}^{\infty} \frac{dt}{t^{1+s}}
\hat{\sum_{m_i,n}}  \exp \left( -\frac{\pi}{V_3 t}
(m_i+ n B^i)^2 -\frac{V_3 \pi}{ t} (n)^2
\right)
\end{equation}
where we have introduced the  singlet charge
 $ n =\frac{1}{3!} \epsilon_3 m^{[3]}$ dual to the three-form charge and
$B^i= \frac{1}{2} \epsilon^{ijk} B_{jk} $ is the dual of the NS 2-form.
We single out the contribution with $n=0$ and for the remaining terms
we Poisson resum on the (unconstrained) integer $m^i$ whose dual charge
is $m_i$. The latter
contribution splits up into a part with $m_i=0$ and a remaining
non-degenerate contribution, so that after some algebra we can write
\begin{multline}
\eis{SO(3,3,\Zint)}{\spi}{s}= \hat{\sum_{m^i} } \left[ \frac{V_3}
{m^i g_{ij}m^j}\right]^{s}
+\frac{2\pi ^{3/2} \Gamma(s-3/2)\zeta(2s-3)V_3^{2-s}}{\Gamma(s)}\\
+\frac{2\pi ^s V_3^{1/2}}{\Gamma(s)}
\hat{\sum_{m_i}}
\hat{\sum_{n}}
\left( \frac{n^2}{m_i g^{ij} m_j} \right)^{\frac{3-2s}{4}}
K_{s-3/2}(2\pi V_3 |n| \sqrt{ m_i g^{ij} m_j}) e ^{2\pi i n m_i B^i}
\end{multline}
In particular for $s=1$ this becomes
\begin{equation}
\label{lvsp3}
\eis{SO(3,3,\Zint)}{\spi}{s=1}= V_3 \eis{SL(3,\Zint)}{\irrep{3}}{s=1}
+ \frac{\pi^2}{3} V_3
+
\pi \hat{\sum_{m_i,n}}
\frac{\exp(-2\pi V_3 |n| \sqrt{ m_i g^{ij} m_j} + 2\pi i n m_i B^i)}
{\sqrt{ m_i g^{ij} m_j}}
\end{equation}
where we have used the definition \eqref{slef}
of the $SL(d)$ Eisenstein series, \eqref{zetav} and \eqref{bessel1}
in each of the three terms
respectively. The two leading terms establish the claim in \eqref{largevolS},
reproducing the trivial and degenerate orbit contribution of
the 1-loop integral $I_3$ respectively. Moreover,
 exact agreement is also explicitly seen \cite{Kiritsis:1997em}
between the third term
and the non-degenerate orbit contribution \eqref{nondeg}.

For the conjugate spinor of $SO(3,3)$ the integral representation
of the Eisenstein series is
\begin{equation}
\eis{SO(3,3,\Zint)}{\spb}{s}
=\frac{\pi ^s}{\Gamma(s)}
\int_{0}^{\infty} \frac{dt}{t^{1+s}}
\hat{\sum_{m_i,n} }  \exp \left( -\frac{\pi}{V_3 t}
(n + m_i B^i)^2
-\frac{V_3 \pi}{ t} m_i g^{ij} m_j\right)
\end{equation}
where in this case we have dualized the two-form into a one-form
$n_1 = \epsilon_3 m^{[2]}/2$, and the dual $B$-field is as above.
In this case, we first separate the $m_i=0$ contributions and for
the remainder Poisson resum on the unconstrained integer $n$, whereafter
we distinguish between $n=0$ and the rest.
After some algebra we then have
\begin{multline}
\eis{SO(3,3,\Zint)}{\spb}{s}= 2\zeta(2s) V_3^s +
\frac{\pi ^{2s-2}\Gamma(2-s)}{\Gamma(s)} \hat{\sum_{m^i} } \left[ \frac{V_3}
{m^i g_{ij}m^j}\right]^{2-s}
\\
+\frac{2\pi ^s \sqrt{V_3}}{\Gamma(s)}
\hat{\sum_{m^i} }
\hat{\sum_n }
\left( \frac{n^2}{m_i g^{ij} m_j} \right)^{\frac{2s-1}{4}}
K_{s-1/2}(2\pi V_3 |n| \sqrt{ m_i g^{ij} m_j}) e ^{2\pi i n m_i B^i}
\end{multline}
where we have also used the identity \eqref{slinv} to rewrite the
second term in terms of the fundamental representation of $SL(d)$.
Setting $s=1$ we find exactly the same result \eqref{lvsp3} as
obtained for the spinor representation, with the first two terms
interchanged as noted in \eqref{largevolC}. The equality of the
$d=3$ spinor and conjugate  series for $s=1$ is obvious from the
fact that the two representations have inverse mass matrices in
this case and (since there are no constraints on the charges) can
hence be related by a complete Poisson resummation
\begin{equation}
\eis{SO(3,3,\Zint)}{\spi}{s} = \eis{SO(3,3,\Zint)}{\spb}{2-s}
\end{equation}
Equivalently, this identity follows from \eqref{eisso33} and \eqref{slinv}.

\subsection{$SO(4,4,\Zint)$ spinor and conjugate spinor Eisenstein series}
Moving on to $SO(4,4)$ we remark that from this case on, one needs to
incorporate the non-trivial half-BPS constraints \eqref{dsconstr} and
\eqref{dpconstr}.
The integral representation of the spinor Eisenstein series reads
\begin{equation}
\eis{SO(4,4,\Zint)}{\spi}{s} =\frac{\pi ^s}{\Gamma(s)}
\int_{0}^{\infty} \frac{dt}{t^{1+s}}\int_0^1 d\theta
\hat{\sum_{m^i,n_i}} \exp \left( -\frac{\pi} {V_4 t}
(m^i+B^{ij}n_j)^2 -\frac{\pi}{ t}  V_4  n_i^2  + 2\pi i \theta m^i n_i
\right)
\end{equation}
where we have dualized the three-form into a one-form $n_i = \frac{1}{3!}
\epsilon_{ijkl} m^{jkl}$ and introduced the dual $B$-field $B^{ij} = \frac{1}{2}
\epsilon^{ijkl} B_{kl}$. The constraint $m^{[1]} \wedge m^{[3]}=0$ then becomes $m^i n_i =0$
and is incorporated due to the $\theta$ integration.
The evaluation of this integral proceeds in a way analogous to the
$SO(d,d)$ vector case, and omitting the details we record the final
result
\begin{multline}
\label{lvspi4} \eis{SO(4,4,\Zint)}{\spi}{s} =
\hat{\sum_{m^i}} \left[ \frac{V_4}
{m^i g_{ij}m^j}\right]^{s}
+
\frac{V_4^{2-s} \pi^{3/2} \Gamma(s-\frac{3}{2})\zeta(2s-3)}{\Gamma(s)\zeta(2s-2)}
\hat{\sum_{n_i} } \left[ \frac{1}{n_i g^{ij} n_j} \right]^{s-1}
+\\
+\frac{4\sqrt{V_4} \pi^{s}}{\Gamma(s)}
\bar{\sum_{m_i,n_i} } 
\frac{1}{2 \zeta( 2s-2)}\, \eis{SL(2,\Zint)}{\irrep{2}}{ s-1 }
 \left({\scriptsize  \begin{matrix} m\cdot m & 
m \cdot  n\\
 m \cdot n &  n\cdot  n \end{matrix}} \right)
\left[|(m\cdot m)(n\cdot n)-(m\cdot n)^2|\right]
^\frac{2s-3}{4}\\
K_{s-\frac{3}{2}} \left(2\pi
V_4 \sqrt{| (m\cdot m)(n\cdot n)-(m\cdot n)^2|} \right)
e ^{2\pi i B^{ij} m_i n_j }
\end{multline}
where all inner products are taken with the inverse metric, and the sum in the last term
runs over non-degenerate $SL(2,\Zint)$ orbits.
In fact, this result can be obtained immediately from the result of
the $SO(4,4)$ vector representation (substitute $d=4$ in \eqref{jv1}
+ \eqref{jv2a} + \eqref{jv2b2}), using the triality relation
\begin{equation}
\M^2 (\spi;g_{ij},B_{ij};m^i,n_i) = \M^2(\vect;V_4 g^{ij}, B^{ij} ;n_i,m^i)
\end{equation}
between the $SO(4,4)$ spinor and vector mass formulae.
For use below we also note that the first two terms can be
expressed in terms of $SL(4)$ Eisenstein
series,
\begin{equation}
\eis{SO(4,4,\Zint)}{\spi}{s} =
V_4^s \eis{SL(4,\Zint)}{\irrep{4}}{s}
+
\frac{V_4^{2-s} \pi^{3/2} \Gamma(s-\frac{3}{2})\zeta(2s-3)}{\Gamma(s)\zeta(2s-2)}
\eis{SL(4,\Zint)}{\irrep{\bar{4}}}{1-s} + \ldots
\end{equation}
Evaluating this at $s=1$ with the use of \eqref{zetav} we reproduce the two leading terms \eqref{largevolS}. Using \eqref{bessel}
the non-degenerate contribution at $s=1$ in \eqref{lvspi4} takes the form
\begin{equation}
2\pi
\bar{\sum_{m_i,n_i} } \frac{
\exp\left(-2\pi V_4 \sqrt{|(m\cdot m)(n\cdot n)-(m\cdot n)^2|}
+ 2\pi i B^{ij} m_i n_j \right) }
{\sqrt{|(m\cdot m)(n\cdot n)-(m\cdot n)^2}|}
\end{equation}
Although we have not been able to show it explicitly, this contribution
should be equal the corresponding non-degenerate contribution in \eqref{lvV}
for
the $SO(4,4)$ vector Eisenstein series at $s=1$, and hence equal
to the non-degenerate contribution of the 1-loop integral $I_4$.

\subsection{$SO(d,d,\Zint)$ spinor and conjugate spinor Eisenstein series}
More generally, we can compute for all $n$ the leading term for the
spinor Eisenstein series, obtained by setting all charges $m^{[3]}=m^{[5]} =
\ldots =0$ except $m^{[1]}$, so that the constraints are trivial. This shows that
\begin{equation}
\eis{SO(d,d,\Zint)}{\spi}{s}
= \hat{\sum_{m^i}} \left[ \frac{V_d}{m^i g_{ij} m^j}
\right]^s + \ldots = V_d^s \eis{SL(d,\Zint)}{\irrep{d}}{s} + \ldots
\end{equation}
so that, for $s=1$, we observe the leading term in \eqref{largevolS}.

For the conjugate spinor, we can go even further and obtain the
first two leading terms. Focusing on the contributions from $m$
and $m^{[2]}$ only and setting $m^{[4]} = m^{[6]} = \ldots =0$ (so that the
constraints can be ignored) we find that
\begin{multline}
\label{sonc}
\eis{SO(d,d,\Zint)}{\spb}{s} = \hat{\sum_{m}} \left[ \frac{V_d}{m^2}
\right]^s +
\frac{\pi^s \Gamma(s-\frac{1}{2})  } { \pi^{s-\frac{1}{2}} \Gamma(s) }
V_d^s \hat{\sum_{m^{ij}} } \left[ \frac{1}{m^{ij} g_{ik} g_{jl} m^{kl} }
\right]^{s-\frac{1}{2}} \delta(m^{[2]}\wedge m^{[2]})+ \ldots
\\
=2 V_d^s \zeta (2s) +
\frac{\pi^s \Gamma(s-\frac{1}{2} ) } { \pi^{s-\frac{1}{2}} \Gamma(s) }
V_d^s \eis{SL(d,\Zint)}{\irrep{[2]}}{s-\frac{1}{2} }
  + \ldots
\end{multline}
Here, the leading term is obtained from $m^{[2]}=0$, while the second
term follows after Poisson resummation on the unconstrained $m$ in
the remainder and setting (the dual) $m=0$. Substituting $s=1$ we
immediately recognize the leading term $\frac{\pi^2}{3}V_d$.

\addcontentsline{toc}{section}{References}
\providecommand{\href}[2]{#2}\begingroup\raggedright

\endgroup
\end{document}